\documentclass[twocolumn]{aastex631}
\usepackage{longtable}
\usepackage{epstopdf}
\usepackage{booktabs}
\usepackage{appendix}

\newcommand{\alf}{Alfv$\acute{\text{e}}$n} 

\shorttitle{Decayless kink oscillations in active region short loops}
\shortauthors{Shrivastav et al. 2024}
\graphicspath{{./}{figures/}}
\begin{document}

\title{On the Existence of Long-Period Decayless Oscillations in Short Active Region Loops}

\author[0000-0001-9035-3245]{Arpit Kumar Shrivastav}
\affiliation{Aryabhatta Research Institute of Observational Sciences, Nainital, India-263002}
\affiliation{Department of Physics, Indian Institute of Science, Bangalore, India-560012}
\author[0000-0002-6954-2276]{Vaibhav Pant}
\affiliation{Aryabhatta Research Institute of Observational Sciences, Nainital, India-263002}
\author[0009-0000-2843-9865]{Rohan Kumar}
\affiliation{Department of Physical Sciences, Indian Institute of Science Education and Research Kolkata, India-700064}
\author[0000-0003-4052-9462]{David Berghmans}
\affiliation{Solar-Terrestrial Centre of Excellence - SIDC, Royal Observatory of Belgium, Ringlaan - 3 - Av. Circulaire, 1180 Brussels, Belgium}
\author[0000-0001-9628-4113]{Tom Van Doorsselaere}
\affiliation{Centre for Mathematical Plasma Astrophysics, Mathematics Department, KU Leuven, Celestijnenlaan 200B bus 2400, B-3001 Leuven, Belgium}
\author[0000-0003-4653-6823]{Dipankar Banerjee}
\affiliation{Aryabhatta Research Institute of Observational Sciences, Nainital, India-263002}
\affiliation{Center of Excellence in Space Science, IISER Kolkata, Kolkata, India-700064}
\author[0000-0002-0175-7449]{Elena Petrova}
\affiliation{Centre for Mathematical Plasma Astrophysics, Mathematics Department, KU Leuven, Celestijnenlaan 200B bus 2400, B-3001 Leuven, Belgium}
\author[0000-0001-9914-9080]{Daye Lim}
\affiliation{Centre for Mathematical Plasma Astrophysics, Mathematics Department, KU Leuven, Celestijnenlaan 200B bus 2400, B-3001 Leuven, Belgium}
\affiliation{Solar-Terrestrial Centre of Excellence - SIDC, Royal Observatory of Belgium, Ringlaan - 3 - Av. Circulaire, 1180 Brussels, Belgium}

\begin{abstract}

Decayless kink oscillations, characterized by their lack of decay in amplitude, have been detected in coronal loops of varying scales in active regions, quiet Sun and coronal holes. Short-period ($<$ 50 s) decayless oscillations have been detected in short loops ($<$ 50 Mm) within active regions. Nevertheless, long-period decayless oscillations in these loops remain relatively unexplored and crucial for understanding the wave modes and excitation mechanisms of decayless oscillations. We present the statistical analysis of decayless oscillations from two active regions observed by the Extreme Ultraviolet Imager (EUI) onboard Solar Orbiter. The average loop length and period of the detected oscillations are 19 Mm and 151 seconds, respectively. We find 82 long-period and 23 short-period oscillations in these loops. We do not obtain a significant correlation between loop length and period. We discuss the possibility of different wave modes in short loops, although standing waves can not be excluded from possible wave modes. Furthermore, a different branch exists for active region short loops in the loop length vs period relation, similar to decayless waves in short loops in quiet Sun and coronal holes. The magnetic fields derived from MHD seismology, based on standing kink modes, show lower values for multiple oscillations compared to previous estimates for long loops in active regions. Additionally, the comparison of period distributions in short loops across different coronal regions indicates that different excitation mechanisms may trigger short-period kink oscillations in active regions compared to the quiet Sun and coronal holes.

\end{abstract}

\keywords{magnetohydrodynamics (MHD) – Sun: corona – Sun: magnetic fields – Sun: oscillations}

\section{Introduction} \label{sec:intro}

\begin{table*}[!ht]
\label{table1}
\caption{Datasets used in this study.}  

\begin{center}
\centering
   \begin{tabular} {@{}cccccccc@{}}

         \hline

  \rule{0pt}{3ex} Date & Time Interval of & Distance from  & Stonyhurst heliographic  & Plate & Cadence  & Field of   \\
  \rule{0pt}{3ex}  & observation (UT)  & the Sun (a.u) & longitude (deg) & scale (km) & (s) &  view (Mm$^{2}$)   \\

     \hline
      \rule{0pt}{3ex} 2022-03-17  & 03:18 - 04:01 & 0.38 & 26.5 & 135 & 3  & 277$\times$277 \\ 
       \rule{0pt}{3ex} 2022-10-13  & 13:06 - 14:35 & 0.29 & -108.7 & 105 & 3 & 215$\times$215 \\   
       
  \hline
\end{tabular}
\end{center}
\end{table*}

\begin{figure*}[!ht]
\centering
\includegraphics[width=0.9\textwidth,clip,trim=0.2cm 1cm 0.2cm 2.5cm]{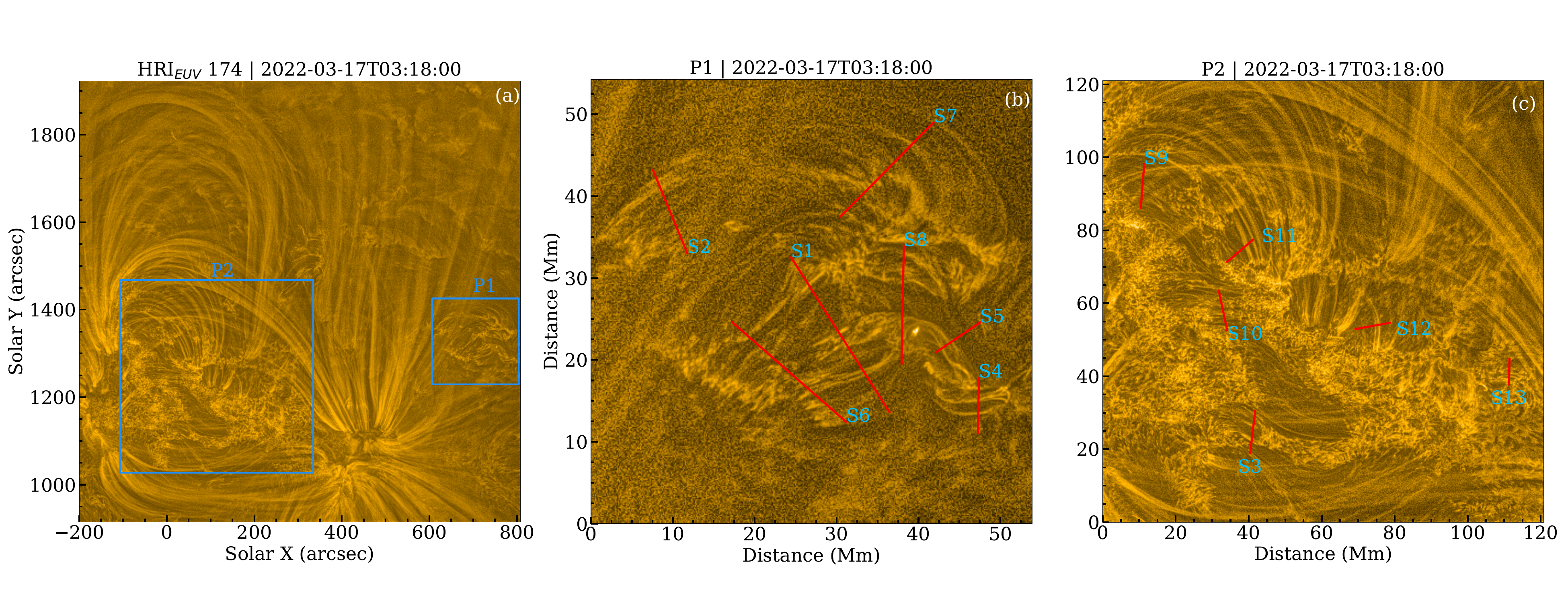}
\includegraphics[width=\textwidth,clip,trim=0.2cm 4.4cm 0.2cm 4.4cm]{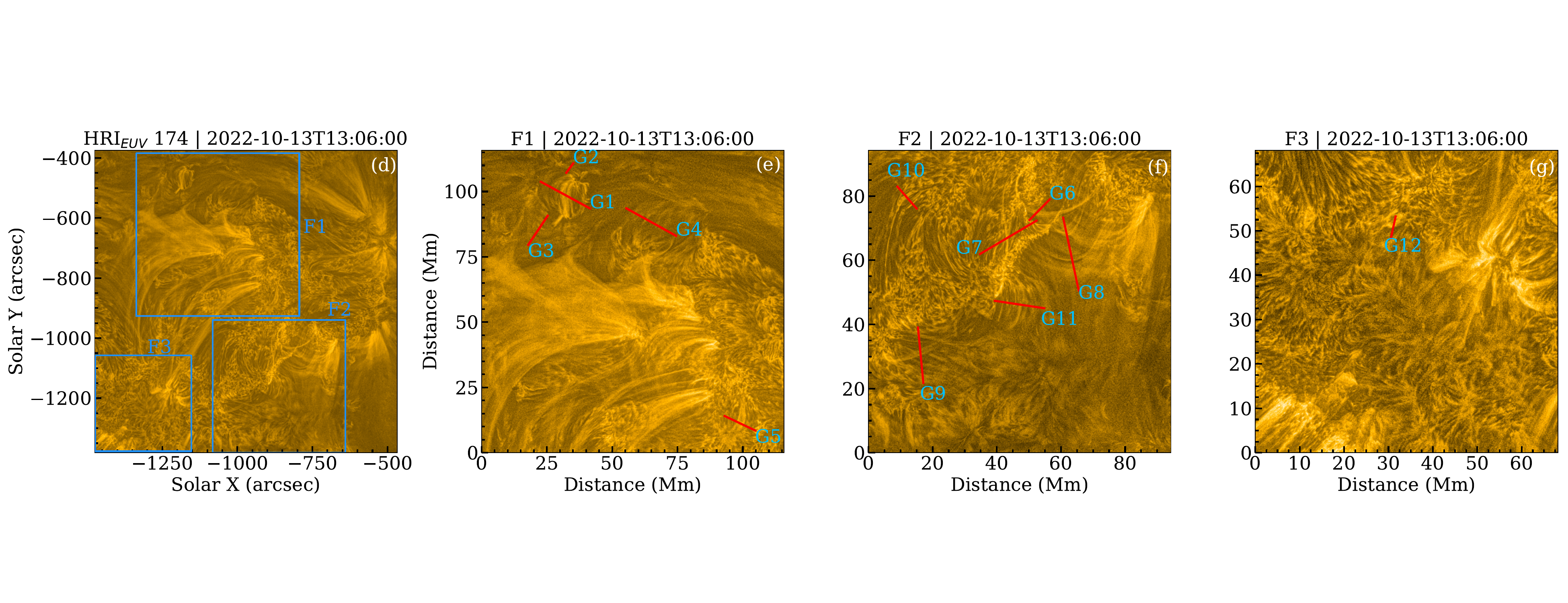}
\caption{Context images of active regions from the datasets. Panels (a) and (d) illustrate the FOV covered by the observations. Within these FOVs, smaller regions have been selected for more detailed analysis, as denoted by the blue boxes. Panels (b)-(c) and (e)-(g) show these smaller ROIs, with the positions of artificial slits indicated by red lines.}
\label{fig:context_d2}
\end{figure*}

 Coronal loops are often observed to show periodic transverse displacement of their axis, which are interpreted as fast kink waves \citep{1999Nakariakov, 1999Schrijver}. The kink waves are one of the normal modes of magnetohydrodynamic (MHD) waves in magnetic cylinders \citep{1983Edwin&Roberts}. The decayless kink oscillations represent a category of kink waves that are omnipresent in the solar corona, persisting for several cycles without any decay in the amplitude \citep{2012Tian, 2013Anfinogentov,2015Anfinogentov,2018Duckenfield,2022Mandal, 2023Li,2024Lim}. These are low-amplitude ($<$ 1Mm) oscillations, and the input energy, ensuring the persistent behaviour, is suggested to be provided from footpoint driving \citep{2016Nakariakov, 2017Karampelas, 2019Guo, 2020Karampelas, 2020Afanasyev, 2021Shi, 2021Ruderman}. Consequently, decayless oscillations are supposed to supply energy from the footpoint to the corona continuously \citep{2023Li_cont}. These characteristics of decayless oscillations position them as one of the possible mechanisms for coronal heating \citep{ 2020tom, 2021NakariakovSSRv, 2023Yuan_natas}.

The recent advancement of imaging instruments accelerated the studies of kink oscillations in the corona, revealing their crucial properties. Atmospheric Imaging Assembly (AIA; \citealt{2012Lemen}) observations indicated a linear relationship between loop length and period for decaying and decayless oscillations with significant correlation coefficients \citep{2015Anfinogentov, 2016Goddard, 2019Nechaeva,2023Zhong}. This confirms the standing nature of kink waves in magnetic flux tubes. AIA observations have been used to investigate the kink oscillation in loops with loop lengths of hundreds of Mm \citep{2013Nistico, 2015Anfinogentov, 2019Anfinogento}. Furthermore, decayless oscillations are also observed to be associated with coronal rains occurring in long active region loops \citep{2024Shrivastav_coronal_rain}.  The limited resolution of AIA poses challenges in resolving short loops.  However, \cite {2022Gao} analyzed decayless kink oscillations in coronal bright points (loop lengths $\sim$23 Mm) in AIA images using motion magnification \citep{2016Anfinogentov, 2021Zhong_motion}.

 The spatial resolution of Solar orbiter's Extreme Ultraviolet Imager \citep[EUI;][]{2020rochus} can reach up to $\sim$200 km near perihelion. This enables the analysis of small-scale structures in the solar corona. 
 The detection of high-frequency kink waves using the EUI in quiet Sun indicated that a large amount of energy may be stored in short loops, sufficient for compensating the radiative losses in the corona \citep{2023Petrova}. Furthermore, the estimation of energy flux of decayless oscillations in short loops rooted in coronal holes and quiet Sun suggests that although high-frequency waves are present, they may not be prevalent in these regions \citep{2023Shrivastav}.  Nonetheless, high-frequency decayless transverse oscillations can play a significant role in coronal heating \citep{2023lim}.

The fine structures in active region moss are seen to exhibit transverse motion \citep{2014Morton, 2015Pant}. Furthermore, \cite{2013Morton_McLaughlin}  showed the presence of propagating waves in these structures. Additionally, the short loops in the active region display the presence of short-period decayless oscillations \citep{2023Li}. The period and loop length of these observed oscillations showed a strong correlation coefficient of 0.98, indicating the standing nature of these short-period waves. However, \cite{2022Gao} and \cite{2023Shrivastav}, in their respective investigations, did not find any significant correlation between the loop length and the oscillation period. As a result, the precise wave mode of these oscillations remains uncertain.  Furthermore, the study of \cite{2023Li} does not include the oscillations with periods exceeding 200 seconds, which may impact the correlation between loop length and period.

In this paper, we analyze short loops in two active regions using EUI, including oscillations with periods greater than 200 s. We report the statistical properties of kink oscillations in active region loops. The paper is arranged as follows: Section 2 explains the observation used for statistical analysis. Section 3 outlines the methodology. Section 4 presents the results and discusses its implications. Following this, the paper concludes with a summary.

\begin{figure*}[!ht]
\centering
\includegraphics[width=0.9\textwidth,clip,trim=0cm 0cm 0cm 0cm]{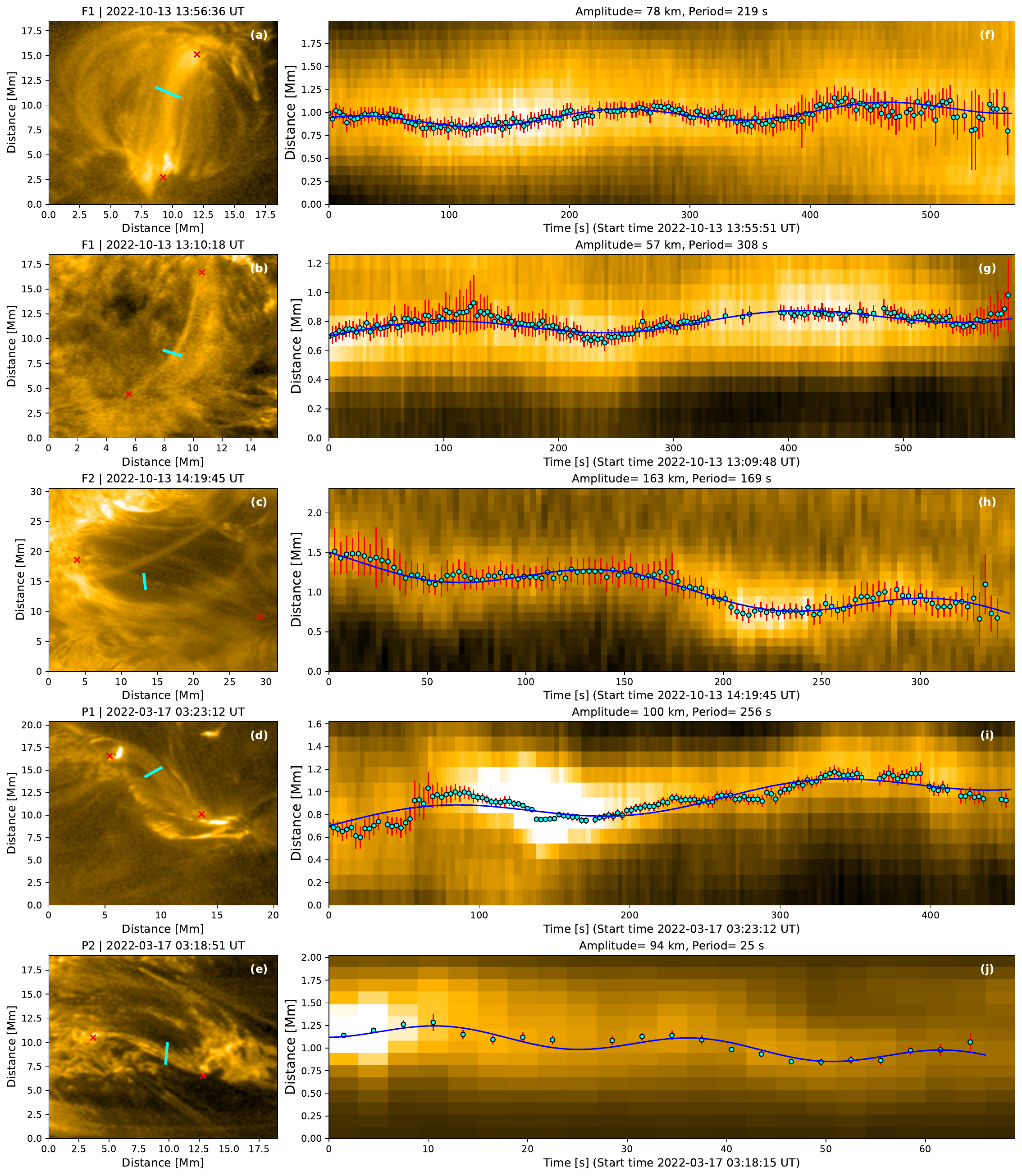}
\caption{The left panels show the loops used for the analysis of oscillations. The cyan lines indicate the artificial slit near the apex, and the red crosses represent the footpoint locations. The right panels show the generated $x-t$ maps from artificial slits. Cyan points represent the position of the loop at any instance, and error bars are provided in red. Blue curves indicate the fitted oscillations. The amplitude and period of the oscillations are provided in the right panels.}
\label{fig:xt-maps}
\end{figure*}

\begin{figure*}[!ht]
\centering
\includegraphics[width=0.9\textwidth,clip,trim=1.8cm 1.1cm 2cm 1.5cm]{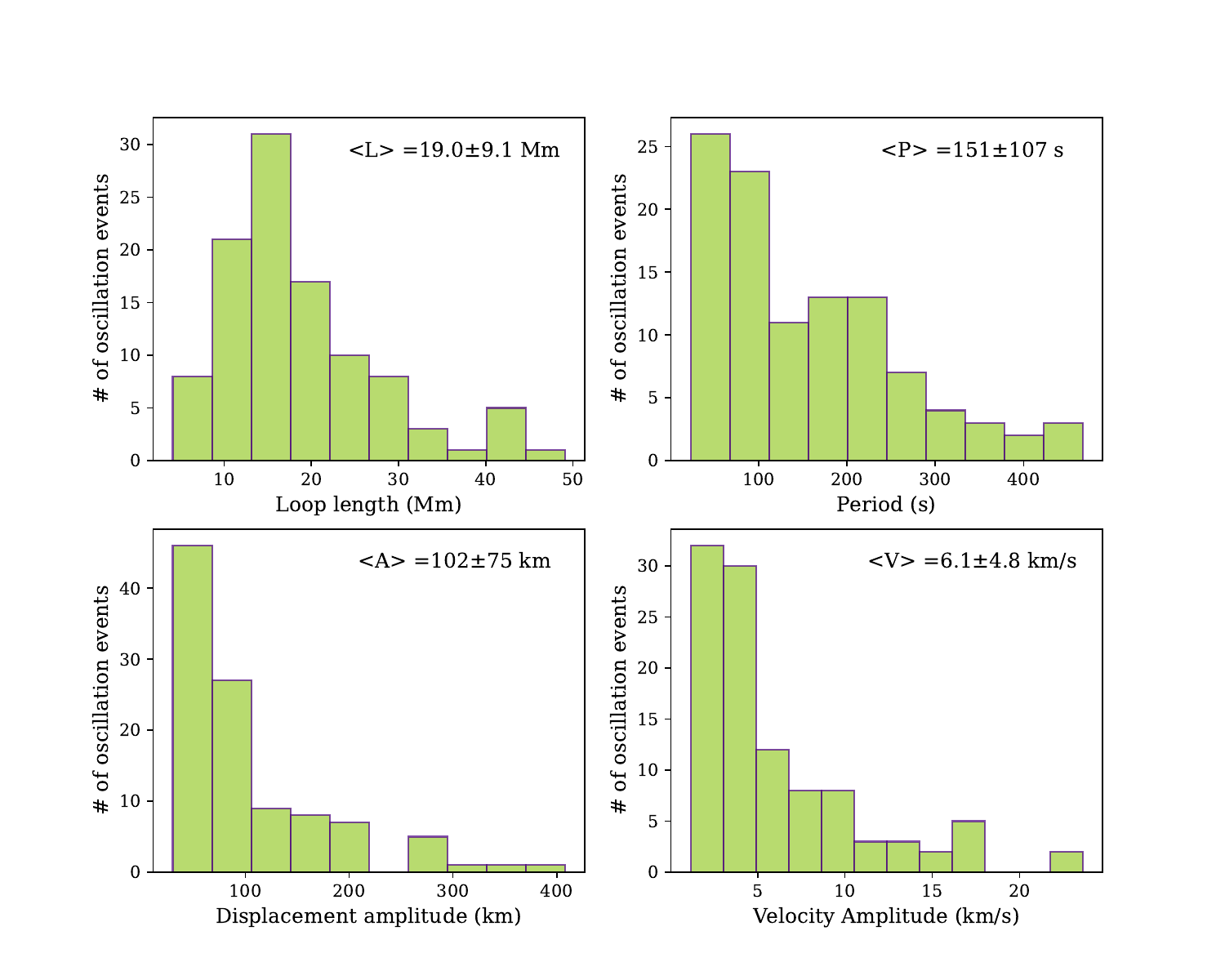}
\caption{Histograms display the distribution of loop oscillation parameters, including the loop length (L), period (P), displacement amplitude (A), and velocity amplitude (V). The average, along with the standard deviation of these distributions, are provided in the figures.  }
\label{fig:parameter_statitics}
\end{figure*}

\section{Details of Observations} \label{sec:obser}

The observational datasets are obtained from High Resolution Imager ($\text{HRI}_{\text{EUV}}$) of the Extreme Ultraviolet Imager telescope onboard Solar Orbiter \citep{2020Mulller}. $\text{HRI}_{\text{EUV}}$ has a field of view (FOV) of about $16.8' \times 16.8'$ and observes the Sun at 174 \AA, dominated by the coronal plasma of $\sim 1\text{ MK}$ attributed to FeIX and FeX emissions. The Level 2 data products are downloaded from the EUI data release 6.0 \citep{euidatarelease6}. We co-align and remove the telescope jitter from the image sequence using the cross-correlation technique \citep{2022Mandal}. The specifications of datasets used for analysis are provided in Table \ref{table1}. The first dataset is the same as employed in \cite{2023Li}, where the active region was the region of interest (ROI).

Figures \ref{fig:context_d2}(a) and (d) display the context image for the datasets, concentrating on active regions. For subsequent analysis, two regions (P1 and P2) from the first dataset and three regions (F1, F2, and F3) from the second dataset are cropped and highlighted by blue boxes. Numerous short loops can be observed in these areas, and the positions of artificial slits for capturing kink oscillation are marked by red lines (see Panels (b)-(c) and (e)-(g) in Figure \ref{fig:context_d2}).

\section{Methodology} \label{sec:analysis}

\begin{figure*}[!ht]
\centering
\includegraphics[width=\textwidth,clip,trim=2.5cm 1.7cm 2.5cm 2.5cm]{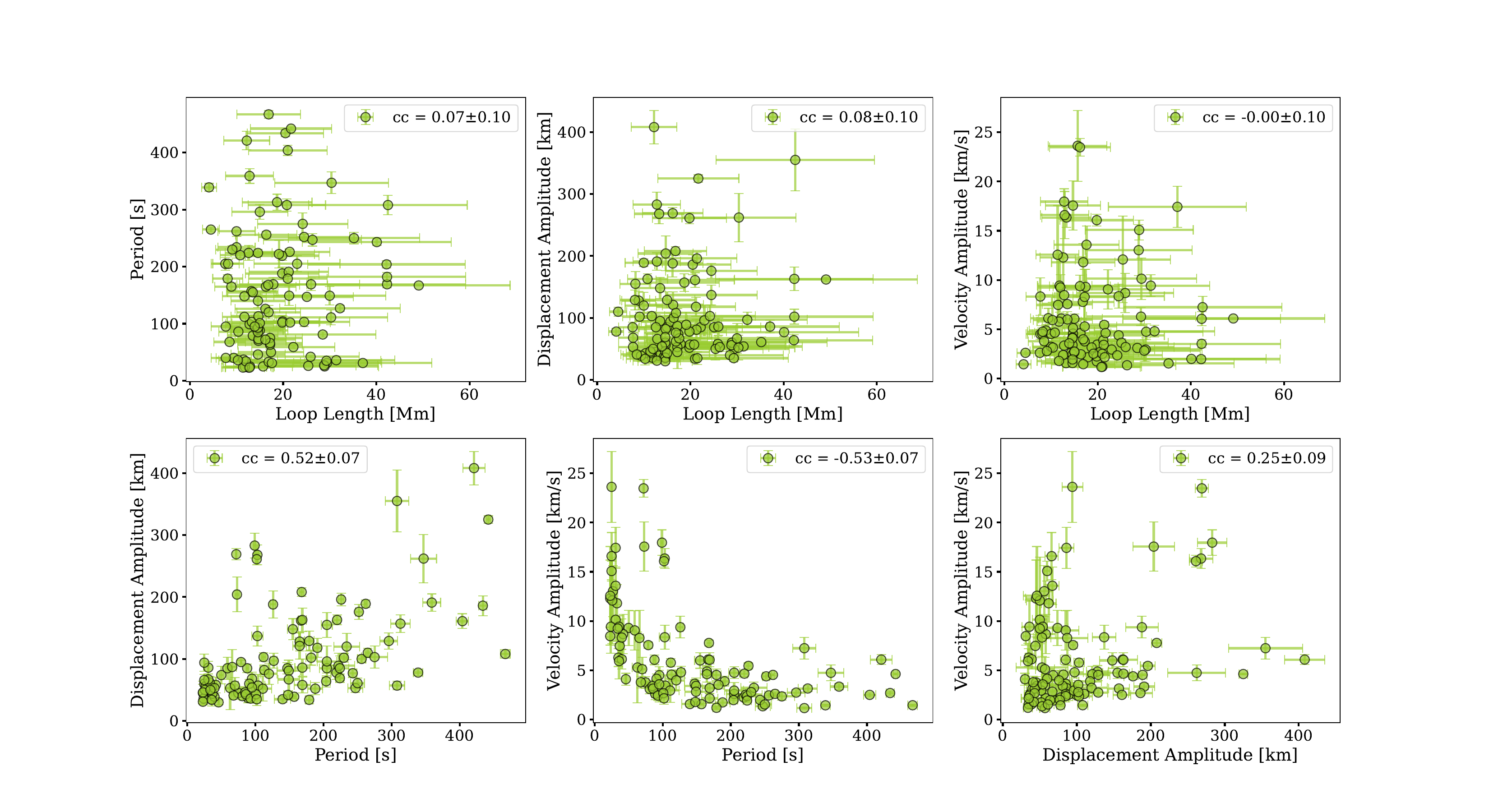}
\caption{Scatter plots illustrate the relation between various loop oscillation parameters. Different pairs of these parameters are used to calculate the linear Pearson correlation coefficients, which are depicted in the plots. The plots also display the estimated standard errors in the correlation coefficient. }
\label{fig:correlation_statistics}
\end{figure*}

Figure \ref{fig:context_d2} shows the presence of several short loops in the active regions observed in the datasets.  Figure \ref{fig:xt-maps}(a)-(f) shows a few examples of short loops selected in this study. The cyan slits depicted in Figure \ref{fig:xt-maps}(a)-(e) have a length sufficient to cover a single loop, and they represent smaller segments of the larger red artificial slits shown in Figure \ref{fig:context_d2}. The slit width is taken to be 5 pixels to increase signal-to-noise ratio \citep{2013Nistico,2015Anfinogentov,2022Gao}. The slit positions for the first dataset are approximately similar to the slit locations used in the study of \cite{2023Li}. In the second dataset, we placed the slits at several loops during the time interval of the dataset and investigated the oscillations regardless of their background complexity. It is important to note that the images presented in Figure \ref{fig:context_d2} have been processed using multi-Gaussian normalization \citep{Morgan&miller2014}, whereas intensity images without any image processing are used for oscillation detection.

The corresponding $x-t$ maps generated using these slits are shown in Figure \ref{fig:xt-maps}(f)-(j). The $x-t$ maps show the presence of kink oscillations. 60\% of the oscillations persist for more than 2 cycles, whereas other oscillations exhibit cycles ranging between 1 and 2. These loops are highly dynamic, overlapping temporally, and lack stability over an extended duration, making it challenging to identify multiple cycles of oscillations \citep{2024Gao, 2023Shrivastav}. However, oscillations with cycles greater than two do not display any notable decay in amplitude over time.

We first estimate the centre of the oscillating loop by fitting a Gaussian function in the slit direction for each time instance. We calculate the uncertainty in the intensity values using the relations provided in \cite{2023Petrova} and \cite{2023Shrivastav}. The uncertainties in the intensity are incorporated while finding the centre of the loop. The error bars on the centre position, shown in Figure \ref{fig:xt-maps}(f)-(j), are standard errors obtained after fitting the Gaussian function. The position of centres is then fitted using the sinusoidal curve with a linear trend of the following form:

\begin{equation}
\label{eq1}
\centering
    A_m(t) = A_{0}+A \sin(\frac{2\pi t}{P}+\phi)+kt.
\end{equation}

where $A$ denotes the amplitude of the oscillation, $P$ is the period of the oscillation, $\phi$ indicates the initial phase of the oscillation, and $k$ estimates the slope of the linear trend. The missing loop positions in Figure \ref{fig:xt-maps}(g) are a result of the failure of Gaussian fitting to converge in certain columns, likely due to a dynamic background or the presence of nearby loops. However, the number of columns with missing positions is minimal compared to the total detected loop positions in the oscillations, so the estimates of period and amplitude should not significantly deviate from the true values. Moreover, such cases are rare in the overall sample. We analyzed several loops and fitted 66 oscillations in the first dataset and 39 oscillations in the second dataset, totalling 105.   We estimate the velocity amplitude of these oscillations by the relation, $V = 2\pi A/P$. The error in the velocity amplitudes is calculated as,  $\sigma_{V}^2 = \left(\frac{\partial V}{\partial P}\sigma_{P} \right)^{2} + \left(\frac{\partial V}{\partial A}\sigma_{A} \right)^{2}$. We compute the loop length, L, by identifying the location of footpoints manually and assuming a three-dimensional semicircular shape of the loop. If R is half of the distance between two footpoints, then, loop length, $L = \pi R$. The red crosses in Figure \ref{fig:xt-maps} (a)-(f) display the approximate positions of loop footpoints. The error in the estimation of loop length is assumed to be $\sim$ 40\% of the loop length \citep{2023Shrivastav}. A few more examples of the fitted oscillations are provided in Appendix \ref{appendix}.

\section{Results \& Discussion} \label{stat_results}

Figure \ref{fig:parameter_statitics} shows the distribution of loop length, period, displacement amplitude, and velocity amplitude for 105 oscillations. The loop length has a range of $\sim$4.1 to 49 Mm, with an average of 19.0$\pm$9.1 Mm. The average loop length is comparable to short loops exhibiting transverse oscillations in active region \citep{2023Li}, quiet Sun \citep{2022Gao, 2023Petrova, 2023Shrivastav} and coronal holes \citep{2023Shrivastav}. The period of these oscillations has a range of $\sim$ 23 to 467 s, with an average value of 151$\pm$107 s. \cite{2022Gao} found a range of 61 to 498 s for transverse oscillations in CBPs rooted in quiet Sun.  The quiet Sun and coronal hole short loops show a period range of 28 to 272 seconds \citep{2023Shrivastav}, which is in the range of values found in this study for active regions. The average amplitude of oscillations is 102$\pm$75 km, with a range of 30 to 408 km. These oscillation amplitudes are similar to decayless oscillations detected in several hundred megameter loops in active regions \citep{2015Anfinogentov, 2018Duckenfield, 2022Mandal} as well as short loops in different regions \citep{2022Gao, 2023Petrova, 2023Li, 2023Shrivastav}. The velocity amplitude ranges from $\sim$ 1 to 24 km s$^{-1}$, with an average of 6.1$\pm$4.8 km s$^{-1}$. The average velocity amplitude is close to those found in quiet Sun and coronal holes \citep{2022Gao, 2023Shrivastav}. Estimated loop length and corresponding oscillation parameters are provided in Table \ref{table2}. The table also includes the mean loop width estimated by averaging the width of the fitted Gaussian during the oscillation time interval. The table can be accessed online\footnote{\href{https://github.com/ArpitkShrivastav/Oscillation-Parameters-Short-Loops}{github.com/ArpitkShrivastav/Oscillation-Parameters-Short-Loops}}.

\begin{figure}[!h]
\centering
\includegraphics[width=0.45\textwidth,clip,trim=1.8cm 1.8cm 2.2cm 2.2cm]{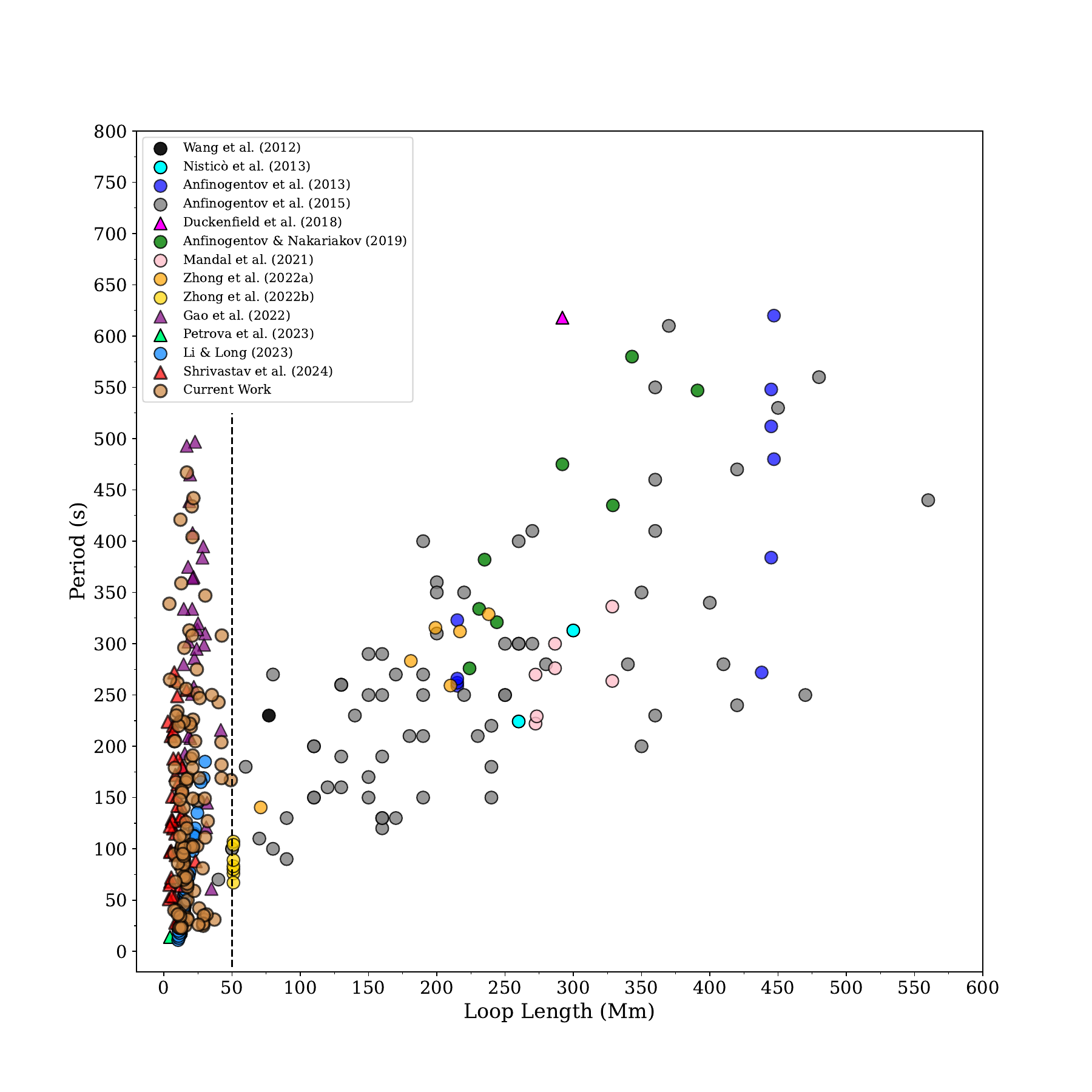}
\caption{Relation between the period and loop length for decayless oscillations, encompassing both long and short loops. These parameters are obtained from previous studies and presented in various colors, while the data points from the current work are specifically displayed in Peru color. The filled circles and triangles represent oscillations found in active regions and quiet Sun regions, respectively.}
\label{fig:all_loop_peri}
\end{figure}

\subsection{Correlation between oscillation parameters}

Figure \ref{fig:correlation_statistics} displays the correlation between the oscillation parameters.  The loop length and period of decayless oscillations in a few hundred Mm loops show a correlation of 0.72 \citep{2015Anfinogentov}. We found a correlation of 0.07$\pm$0.10 between the loop length and oscillation period. No correlation between loop length and period can also indicate the presence of a different wave mode than the standing kink mode \citep{2023Shrivastav}. Figure \ref{fig:all_loop_peri} illustrates the relation between loop length and period for decayless oscillations analyzed in coronal loops. The figure encompasses oscillations from the current study, combined with results from previous studies \citep{2012Wang, 2013Nistico, 2013Anfinogentov, 2015Anfinogentov, 2018Duckenfield, 2019Anfinogento, 2021Mandal_flareoscl, 2022Zhonga, 2023Petrova, 2022Zhong, 2022Gao, 2023Li, 2023Shrivastav}. The oscillations detected in active regions are denoted by filled circles, whereas the oscillations from quiet Sun and coronal holes are indicated by filled triangles.  The figure also suggests the presence of another branch, separate from the one followed by loops with lengths of hundreds of Mm. The dashed line in Figure \ref{fig:all_loop_peri} marks the boundary of this branch at 50 Mm. A significant correlation between loop length and period is observed for oscillations in loops longer than 50 Mm. These are referred to as long loops in previous studies. In this paper, however, we focus only on loops shorter than 50 Mm, which we refer to as short loops. If the trend for long loops had continued for loops shorter than 50 Mm, the periods would be under 50 seconds. So, we separate the periods into two groups: short periods for those under 50 seconds and long periods for those over 50 seconds. The oscillations in the shorter loops in quiet Sun and coronal hole also showed a different branch and slope compared to longer loops in the loop length vs period diagram \citep{2023Shrivastav}. The existence of long-period oscillations in active region short loops indicates that they are following the same branch as quiet Sun and coronal holes (see Figure \ref{fig:all_loop_peri}).

The correlation of loop length with displacement amplitude and velocity amplitude is not significant (Figure \ref{fig:correlation_statistics}), similar to quiet Sun and coronal holes short loops \citep{2022Gao, 2023Shrivastav}. A weaker correlation between period and displacement amplitude is obtained, although it is not significant to conclude any interpretation. The correlation of velocity amplitude with period and displacement amplitude can be affected by the relation between them \citep{2022Gao, 2023Shrivastav}.

\begin{figure}[!ht]
\centering
\includegraphics[width=0.5\textwidth,clip,trim=4cm 0.5cm 3cm 2.5cm]{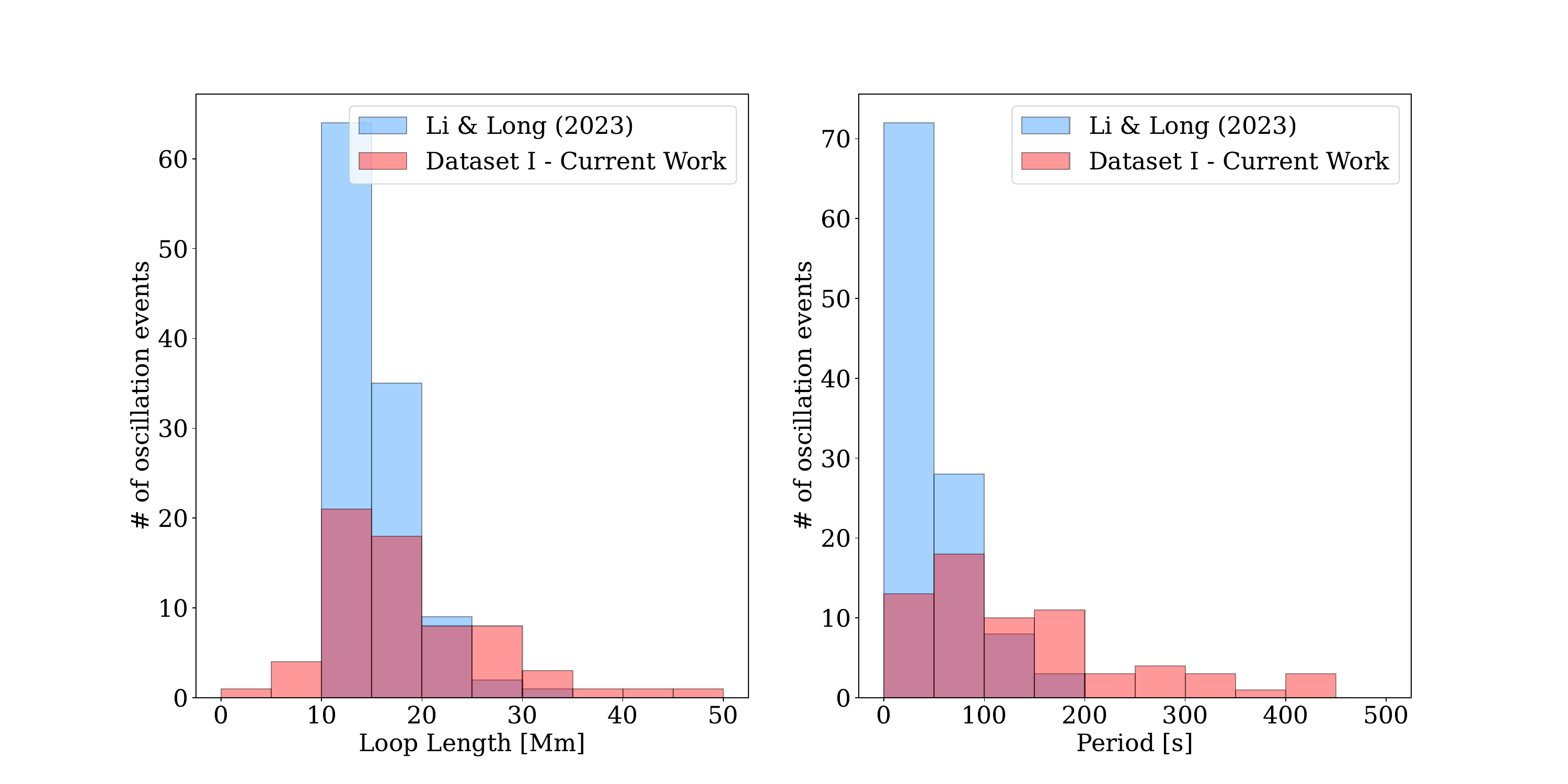}
\caption{The figure shows the distribution of periods and loop lengths obtained in \cite{2023Li} and dataset -I of the current work.}
\label{fig:lilong_compare}
\end{figure}

\subsection{Comparison of oscillation properties with \cite{2023Li}}

Since the first dataset is the same as used in \cite{2023Li} and slit locations are approximately similar, we compare the oscillation properties obtained in the current work with those obtained in \cite{2023Li}. Figure \ref{fig:lilong_compare} compares the distribution of loop lengths and periods estimated in the same dataset of the current work to those of \cite{2023Li}. In our analysis, we identify 66 oscillations in dataset-I, with loop lengths ranging from 4.5 to 49 Mm, having an average of 19.0 Mm. \cite{2023Li} examined oscillations in loops with lengths spanning from 10.5 to 30.2 Mm with an average of 15 Mm, which is similar to loop lengths obtained in this work. The periods in the dataset-I have a range of 23 to 442 s, while \cite{2023Li} observed periods ranging from 11 to 185 s, with an average of 49 seconds. We observe a broader range of periods within these short loops by including the oscillations with periods greater than 200 seconds.  The right panel in Figure \ref{fig:lilong_compare} illustrates that the short-period oscillations in the study of \cite{2023Li} are large compared to the dataset-I in the current work. Several reasons could produce this discrepancy. As mentioned in \cite{2023Li}, the loop centres and edges are manually identified to extract the oscillation properties. However, we adopt the criterion of finding the position of the loop by fitting the Gaussian in the slit direction. This criterion rejected several threads in both datasets in the current study, as the Gaussian fitting was not able to converge at different time frames in $x-t$ maps. This could be due to overlapping structures along the slit. It is also not clear whether those threads truly represent oscillations or if the interaction between loops is causing an apparent effect, and this fact needs further investigation.  Furthermore, the amplitudes obtained in the current work are more accurate because they are estimated after considering associated uncertainties in the intensity values. Additionally, we do not include oscillations with cycles less than 1, which can reduce the number of short-period oscillations in the current work. In a nutshell, the method used in the current work for detecting oscillations is robust and stringent compared to \cite{2023Li}, which can partially account for less high-frequency oscillations in the current study.

\subsection{Wave mode of the oscillations} \label{wave_mode}

\begin{figure}[!ht]
\centering
\includegraphics[width=0.49\textwidth,clip,trim=0cm 0cm 0cm 0cm]{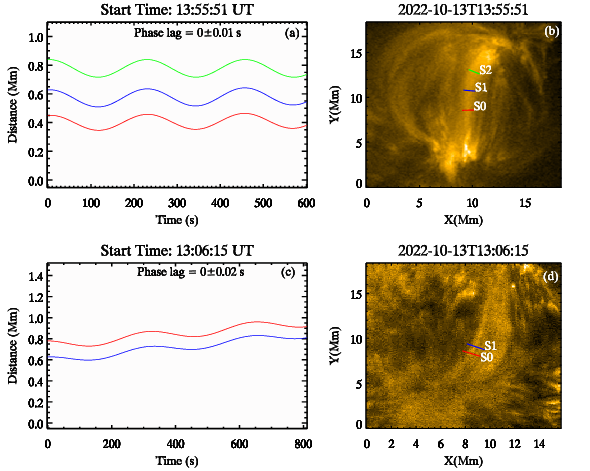}
\caption{{Phase lag analysis of the oscillations. The figure presents the multi-slit analysis for the two loops shown in panels (b) and (d). S0 and S1 are two slits that were placed at different positions of the loops. Panel (b) also include an additional slit, S2, near one footpoint of the loop. Panels (a) and (c) show the fitted oscillation profile from slits indicated in panels (b) and (d). Plots also indicate the phase lag and associated uncertainty.} }
\label{fig:wave_mode}
\end{figure}

Figure \ref{fig:wave_mode} illustrates the results of phase lag for oscillations observed at various slit positions along two loops. Panels (b) and (d) display two slits, S0 and S1, used for phase lag analysis. Prior to phase lag calculation for improved oscillation detection, background subtraction is applied to these $x-t$ maps. The background is constructed by smoothing the $x-t$ maps with a time window longer than oscillation periods. The fitted profiles corresponding to these oscillations are presented in Panels (a) and (c). The phase lag is calculated by finding the time shift of maximum cross-correlation between the detrended oscillation profiles from two loop positions. This results in values of 0$\pm$0.01 s and 0$\pm$0.02 s, respectively. The absence of phase lag between oscillations at different slits may suggest the presence of standing waves within these loops. Similar phase lags have been observed in decayless kink waves in quiet Sun and coronal hole short loops \citep{2023Shrivastav}. However, the lack of a correlation between loop length and period hinders the interpretation of standing modes. Moreover, slit S2 on the first loop is positioned near one footpoint, and the fitted oscillation for this slit exhibits the same displacement amplitude and period as slit S1 which is near the apex of the same loop. However, this characteristic could result from the dynamic nature of short loops or possibly from the influence of the driving source \citep{2023Li_travelling}. Additionally, short loops can lack good contrast with respect to the background along their length, and nearby regions may contain overlapping structures, making it challenging to capture oscillations away from the loop top \citep{2023Shrivastav}. These factors could be possible reasons why a proper oscillation signature away from the apex is not observed in the second loop (Figure \ref{fig:wave_mode} (d)).

The inclined p-mode driver can excite a longer period oscillation in a short coronal loop as reported by \cite{2023Gao} using a numerical simulation. Consequently, the long period of oscillations observed in short loops within active regions in this study may serve as observational support for this phenomenon. These long periods of oscillations might represent the period of the driver, offering a possible explanation for the absence of a correlation between loop length and period. However, the simultaneous detection of the long and short periods in the same loop, as observed in the simulations by \cite{2023Gao}, requires further investigation using different datasets. Furthermore, the vertical displacement as kink-like motions within short coronal loops can arise from the excitation of slow modes by external p-modes \citep{2024Lopin_Nagorny}. The long-period kink waves identified in the current study might also represent driven slow-mode waves. However, these wave modes require strong intensity variations along the loop, which is not observed in the current datasets. The 3D simulation of decayless oscillation from broadband driver suggested the excitation of half harmonic due to the presence of the transition region in the model \citep{2024Karampelas}, which could also be related to long period oscillations observed in the current work. However, periods close to the fundamental mode and half harmonic were found in the single loop, which is not detected in this study and requires future investigation. Additionally, it is worth noting that a fast-propagating wave crossing the slits in Figure \ref{fig:wave_mode} within a time interval shorter than the observation cadence may go undetected, remaining a viable wave mode \citep{2023Shrivastav}. Therefore, several mechanisms can account for the observations presented in this study.

\subsection{Distribution of kink speed and Magnetic field from seismology}

\begin{figure}[!h]
\centering
\includegraphics[width=0.49\textwidth,clip,trim=2.8cm 0.3cm 3cm 1.2cm]{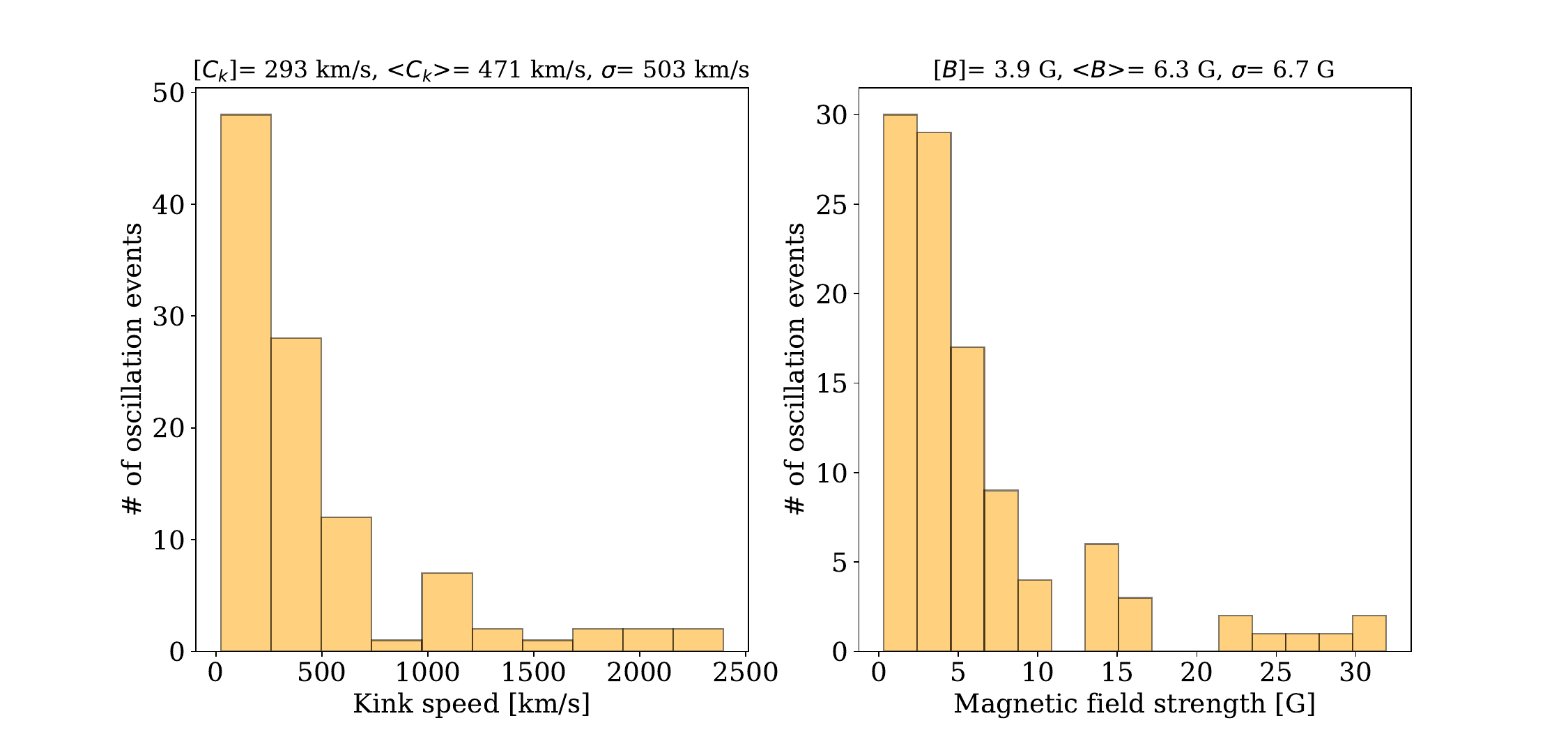}
\caption{The distribution of kink speed derived from coronal seismology is depicted in the left panel, while the right panel shows the histogram of the estimated magnetic field. The figures include the median, mean, and standard deviation values for each distribution. }
\label{fig:coronal_seismology}
\end{figure}

As we discussed, several wave modes with different excitation mechanisms can possibly explain the observed properties. However, since we do not find any phase lag at different positions of loops, standing kink modes can not be ruled out for the observed oscillations. The kink speed for a fundamental standing kink mode can be calculated using the formula,

\begin{equation}
    C_{k} = \frac{2L}{P}.
\end{equation}

 It should be noted that this relation will not be applicable in the case of oscillations not representing a wave mode, as indicated in the previous discussion, which is a caveat of using this relation.  The estimated kink speed ranges from 24 to 2394 km s$^{-1}$ with a mean value of 471 km s$^{-1}$ (Figure \ref{fig:coronal_seismology}). The mean value is comparatively lower than the kink speed found in active region corona \citep{2013Nistico, 2016Goddard, 2019Anfinogento}. Additionally, if these kink oscillations are linked to slow mode waves, it would provide an explanation for the lower kink speed obtained for several oscillations in the current study. The lower value of kink speed could also result from the uncertainty associated with the measurement of loop length \citep{2022Gao, 2023Shrivastav}. The magnetic field can be calculated using the relation, 

\begin{equation}
    B = C_{k}\sqrt{\frac{1+\zeta}{2}}\sqrt{\mu_{0}\rho_{i}\widetilde{m}}, 
\end{equation}

\begin{figure*}[!ht]
\centering
\includegraphics[width=0.45\textwidth,clip,trim=0.2cm 1.1cm 2.2cm 2.2cm]{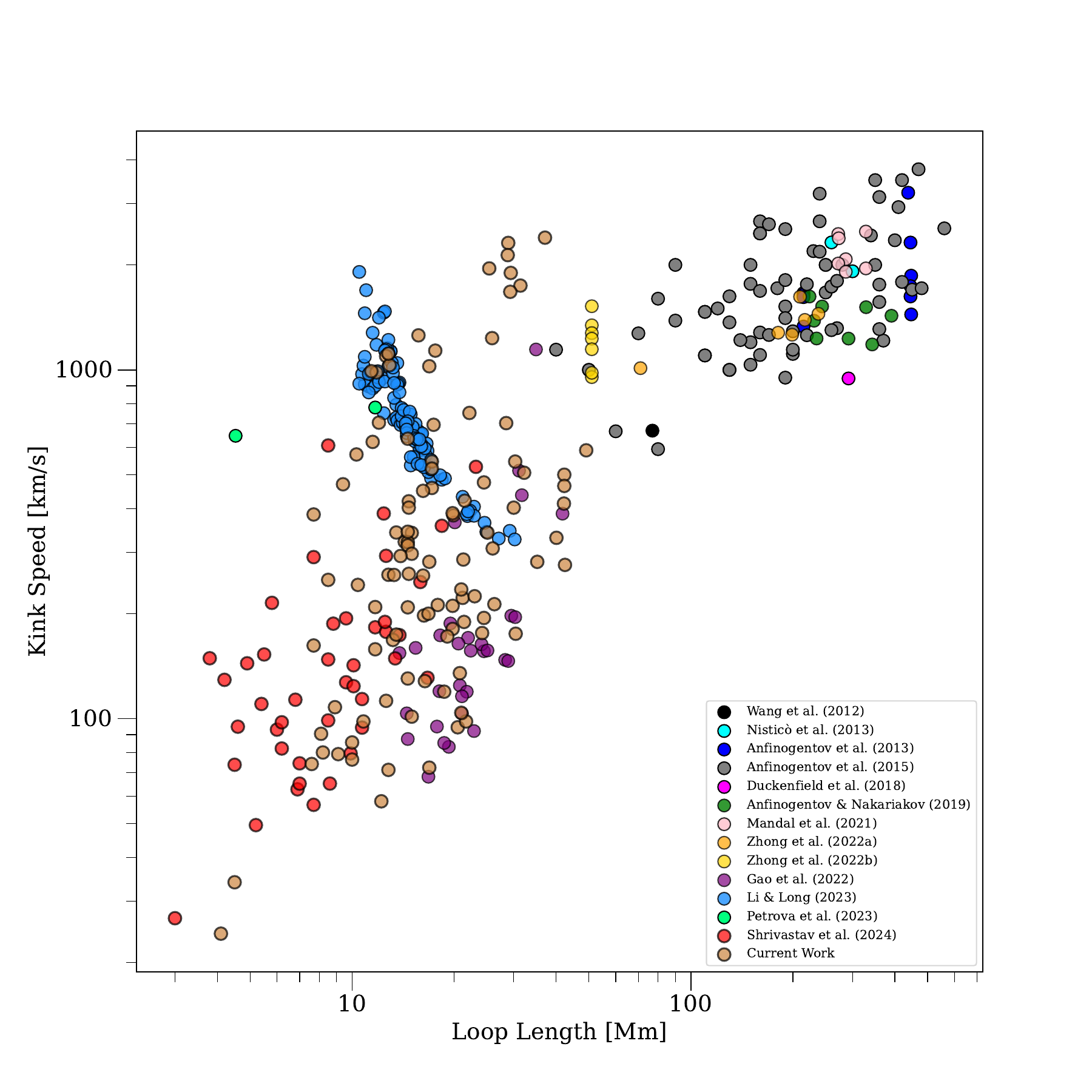}
\includegraphics[width=0.45\textwidth,clip,trim=1cm 1cm 1cm 1cm]{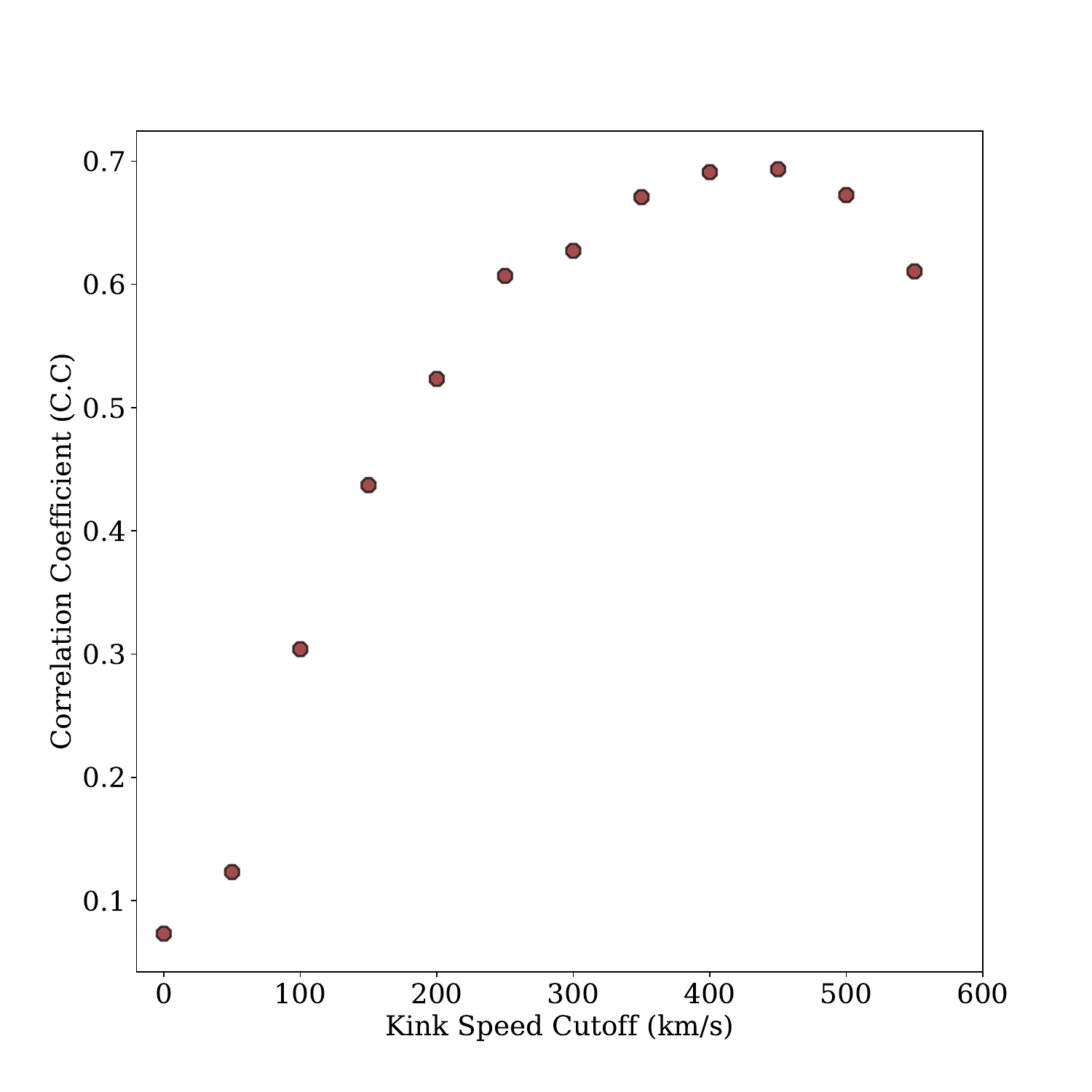}
\caption{The left panel illustrates the relationship between kink speed and loop lengths, collecting findings from previous studies on decayless oscillations. Meanwhile, the right panel demonstrates the variation in the correlation coefficient between loop length and period with the kink speed cutoff.}
\label{fig:ck_vs_L}
\end{figure*}

where $\rho_{i}$ is loop density and $\zeta$ is the density contrast of loop and background. Taking the value of $\zeta$ as 1/3 and a coronal loop density of 1.67$\times$ 10$^{-12}$ kg m$^{-3}$ \citep{2023Petrova}, we calculate the magnetic field in the range of 0.3 to 31 G. The mean value of the magnetic field distribution is determined to be 6.3 G, as shown in the right panel of Figure \ref{fig:coronal_seismology}. Previous magnetic field estimation using coronal seismology in active region loops corresponds to values in the range of a few Gauss to a few tens of Gauss \citep{2001Nakariakov_Ofman, 2013Nistico, 2016Pascoe,2022Zhang}. The average magnetic field obtained here is at the lower end of the range. However, approximately 21\% of oscillations correspond to magnetic field values below 2 G, with 10 oscillations having magnetic field values close to 1 G. These low magnetic field values in active regions can result from waves being driven by a footpoint driver. In instances of such oscillations, the oscillation period might not be dependent on the loop properties but rather on the driver \citep{2022Gao,2023Gao}. In conclusion, the coronal seismology technique should be applied with caution in short loops due to the possibility of multiple wave modes that can be challenging to detect in imaging datasets.

Recent studies have revealed a significant correlation between loop length and kink speed of oscillations observed in short loops \citep{2022Gao, 2024Gao}. This correlation is anticipated due to the increase in \alf\ speeds with height and given that the major radii of these loops span from the chromosphere to the corona. The left panel of Figure \ref{fig:ck_vs_L} illustrates the variation of kink speed with loop length in log-log space, combining our present findings with previous studies on decayless oscillations. Additionally, Figure \ref{fig:ck_vs_L} demonstrates the existence of several oscillations with kink speeds lower than 200 km/s. In the current study, we observe a correlation coefficient of 0.5 between kink speed and loop length in log-log space. This coefficient is lower compared to those found in CBPs \citep{2022Gao} and transition region loops \citep{2024Gao}. However, the sample size used in our present work is extensive and suggests that the kink speed only weakly increases with height. Moreover, the possibility of oscillations in our present work not representing the wave mode could influence the estimation of kink speed and, hence, the correlation coefficient.  

If these short loops have coronal density, the kink speeds of the fundamental mode will be comparable with those of larger loops. To explore this notion, we calculated the correlation coefficient between loop length and period by considering the sample with a kink speed greater than a specified value, termed the kink speed cutoff. Subsequently, we plotted the variation of the correlation coefficient with increasing kink speed cutoff in the right panel of Figure \ref{fig:ck_vs_L}. The correlation coefficient increases as the kink speed cutoff increases and reaches a value of 0.7 (p-value $<$0.05) at a kink speed cutoff of 400 km/s for a sample size of 39 oscillations. The reduction in the correlation coefficient later on can be attributed to the smaller number of samples at those cutoff speeds. \cite{2015Anfinogentov} observed a correlation coefficient of 0.72 between loop length and period for decayless oscillations in several hundred megameter loops. This suggests that oscillations with kink speeds lower than 400 km/s may not represent fundamental kink modes, implying that the reported oscillations in this work could constitute a mixture of different waves discussed in section \ref{wave_mode}. The various possibilities of waves are not distinguishable from the current dataset, and further studies are necessary to confirm or rule out those possibilities. 

\subsection{Period distribution in different regions}

\begin{figure}[!h]
\centering
\includegraphics[width=0.46\textwidth,clip,trim=0cm 0cm 0cm 0cm]{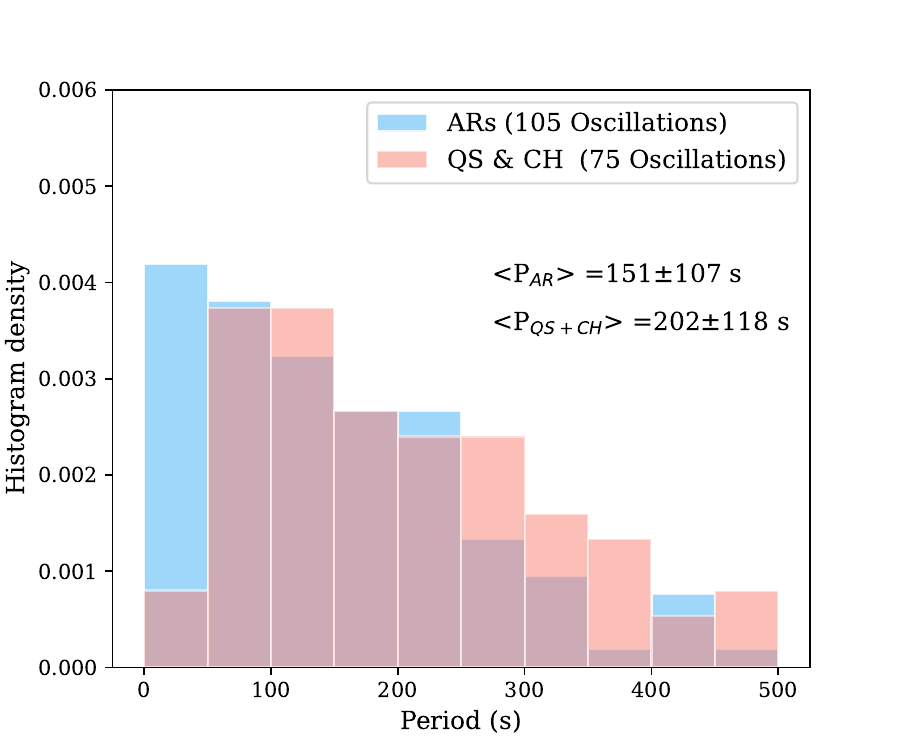}
    \caption{The distribution of oscillation periods in different regions.}
\label{fig:AR_vs_qs}
\end{figure}

We combine the oscillation data derived from studies \cite{2022Gao,2023Petrova} and \cite{2023Shrivastav} to construct the period distribution of short loops in both quiet Sun and coronal holes. In Figure \ref{fig:AR_vs_qs}, the distribution of oscillation periods in short loops from quiet Sun and coronal holes is presented and compared with those observed in active regions in the current study. With 75 oscillations analysed in quiet Sun and coronal holes and 105 from active regions, we normalize the histograms for a comprehensive comparison, plotting histogram density (refer to Figure \ref{fig:AR_vs_qs}). The period distribution in quiet Sun and coronal holes significantly differ from each other in the region of short periods ($<$50 s). The short-period oscillation events in quiet Sun and coronal holes are fewer compared to those in active regions, possibly due to these two reasons. First, \cite{2022Gao} utilized AIA data with a 12-second cadence to investigate kink oscillations in CBPs; hence, the detection of short-period oscillations is limited due to instrument cadence in this particular case. Furthermore, \cite{2023Shrivastav} concluded that short-period oscillations might not be prevalent in short loops within quiet Sun and coronal holes. Therefore, either the actual short-period events are lower in quiet Sun and coronal holes as these regions require lower energy flux than active regions or additional studies on decayless oscillations in these regions using high-cadence data from EUI are necessary to refute this notion. However, the notable difference in the distribution within the short-period regime indicates that different driving mechanisms may trigger short-period oscillations in short loops across different regions. It is crucial to note that the number of oscillations from the quiet Sun is greater than those from coronal holes as \cite{2023Shrivastav} combined the oscillations from the quiet Sun and coronal holes.

\section{Conclusion and Summary} \label{sec:conclusion}

We present a statistical analysis of kink oscillations in short loops in two active regions observed using EUI instruments. Previously studied kink waves in active region short loops were not investigated with periods greater than 200 seconds, which impacts the overall relationship between loop length and period in these loops. The average loop length is estimated to be 19.0$\pm$9.1 Mm. We find the presence of several long-period oscillations in short loops embedded in active regions. Furthermore, the decayless oscillations in active regions display two branches in loop length vs. period relation. The period distribution of kink oscillations in active region short loops differs from those obtained in quiet Sun and coronal holes in the short period regime, suggesting a different driving mechanism for short-period oscillations in different regions on the Sun. The wave modes of these oscillations are uncertain, and distinguishing different wave modes requires further studies of short loops utilizing high-cadence imaging and spectroscopic observations. In this context, numerical simulations of kink waves in short loops will be crucial.   

\begin{acknowledgements}
\textit{Acknowledgements:} A.K.S is supported by funds of the Council of Scientific \& Industrial Research (CSIR), India, under file no. 09/079(2872)/2021-EMR-I. R.K. acknowledge the support of the NIUS programme of HBCSE-TIFR funded by the Department of Atomic Energy, Govt. of India (Project No. RTI4001). V.P. is supported by SERB start-up research grant (File no. SRG/2022/001687). TVD was supported by the C1 grant TRACEspace of Internal Funds KU Leuven, and a Senior Research Project (G088021N) of the FWO Vlaanderen. Furthermore, TVD received financial support from the Flemish Government under the long-term structural Methusalem funding program, project SOUL: Stellar evolution in full glory, grant METH/24/012 at KU Leuven. The research that led to these results was subsidised by the Belgian Federal Science Policy Office through the contract B2/223/P1/CLOSE-UP. This project DynaSun has received funding under the Horizon Europe programme of the European Union under grant agreement (no. 101131534). Views and opinions expressed are however those of the author(s) only and do not necessarily reflect those of the European Union and therefore the European Union cannot be held responsible for them. TVD is grateful for the hospitality of ARIES, Nainital for his visit in spring 2023 where these results were discussed. Solar Orbiter is a space mission of international collaboration between ESA and NASA, operated by ESA. The EUI instrument was built by CSL, IAS, MPS, MSSL/UCL, PMOD/WRC, ROB, LCF/IO with funding from the Belgian Federal Science Policy Office (BELSPO/PRODEX PEA 4000112292 and 4000134088); the Centre National d’Etudes Spatiales (CNES); the UK Space Agency (UKSA); the Bundesministerium f\"{u}r Wirtschaft und Energie (BMWi) through the Deutsches Zentrum f\"{u}r Luft- und Raumfahrt (DLR); and the Swiss Space Office (SSO). 
\end{acknowledgements}

\bibliography{kink_AR_solo}{}

\begin{thebibliography}{}
\expandafter\ifx\csname natexlab\endcsname\relax\def\natexlab#1{#1}\fi
\providecommand{\url}[1]{\href{#1}{#1}}
\providecommand{\dodoi}[1]{doi:~\href{http://doi.org/#1}{\nolinkurl{#1}}}
\providecommand{\doeprint}[1]{\href{http://ascl.net/#1}{\nolinkurl{http://ascl.net/#1}}}
\providecommand{\doarXiv}[1]{\href{https://arxiv.org/abs/#1}{\nolinkurl{https://arxiv.org/abs/#1}}}

\bibitem[{{Afanasyev} {et~al.}(2020){Afanasyev}, {Van Doorsselaere}, \&
  {Nakariakov}}]{2020Afanasyev}
{Afanasyev}, A.~N., {Van Doorsselaere}, T., \& {Nakariakov}, V.~M. 2020, \aap,
  633, L8, \dodoi{10.1051/0004-6361/201937187}

\bibitem[{{Anfinogentov} \& {Nakariakov}(2016)}]{2016Anfinogentov}
{Anfinogentov}, S., \& {Nakariakov}, V.~M. 2016, \solphys, 291, 3251,
  \dodoi{10.1007/s11207-016-1013-z}

\bibitem[{{Anfinogentov} {et~al.}(2013){Anfinogentov}, {Nistic{\`o}}, \&
  {Nakariakov}}]{2013Anfinogentov}
{Anfinogentov}, S., {Nistic{\`o}}, G., \& {Nakariakov}, V.~M. 2013, \aap, 560,
  A107, \dodoi{10.1051/0004-6361/201322094}

\bibitem[{{Anfinogentov} \& {Nakariakov}(2019)}]{2019Anfinogento}
{Anfinogentov}, S.~A., \& {Nakariakov}, V.~M. 2019, \apjl, 884, L40,
  \dodoi{10.3847/2041-8213/ab4792}

\bibitem[{{Anfinogentov} {et~al.}(2015){Anfinogentov}, {Nakariakov}, \&
  {Nistic{\`o}}}]{2015Anfinogentov}
{Anfinogentov}, S.~A., {Nakariakov}, V.~M., \& {Nistic{\`o}}, G. 2015, \aap,
  583, A136, \dodoi{10.1051/0004-6361/201526195}

\bibitem[{{Duckenfield} {et~al.}(2018){Duckenfield}, {Anfinogentov}, {Pascoe},
  \& {Nakariakov}}]{2018Duckenfield}
{Duckenfield}, T., {Anfinogentov}, S.~A., {Pascoe}, D.~J., \& {Nakariakov},
  V.~M. 2018, \apjl, 854, L5, \dodoi{10.3847/2041-8213/aaaaeb}

\bibitem[{{Edwin} \& {Roberts}(1983)}]{1983Edwin&Roberts}
{Edwin}, P.~M., \& {Roberts}, B. 1983, \solphys, 88, 179,
  \dodoi{10.1007/BF00196186}

\bibitem[{{Gao} {et~al.}(2023){Gao}, {Guo}, {Van Doorsselaere}, {Tian}, \&
  {Skirvin}}]{2023Gao}
{Gao}, Y., {Guo}, M., {Van Doorsselaere}, T., {Tian}, H., \& {Skirvin}, S.~J.
  2023, \apj, 955, 73, \dodoi{10.3847/1538-4357/acf454}

\bibitem[{{Gao} {et~al.}(2024){Gao}, {Hou}, {Van Doorsselaere}, \&
  {Guo}}]{2024Gao}
{Gao}, Y., {Hou}, Z., {Van Doorsselaere}, T., \& {Guo}, M. 2024, \aap, 681, L4,
  \dodoi{10.1051/0004-6361/202348702}

\bibitem[{{Gao} {et~al.}(2022){Gao}, {Tian}, {Van Doorsselaere}, \&
  {Chen}}]{2022Gao}
{Gao}, Y., {Tian}, H., {Van Doorsselaere}, T., \& {Chen}, Y. 2022, \apj, 930,
  55, \dodoi{10.3847/1538-4357/ac62cf}

\bibitem[{{Goddard} {et~al.}(2016){Goddard}, {Nistic{\`o}}, {Nakariakov}, \&
  {Zimovets}}]{2016Goddard}
{Goddard}, C.~R., {Nistic{\`o}}, G., {Nakariakov}, V.~M., \& {Zimovets}, I.~V.
  2016, \aap, 585, A137, \dodoi{10.1051/0004-6361/201527341}

\bibitem[{{Guo} {et~al.}(2019){Guo}, {Van Doorsselaere}, {Karampelas}, {Li},
  {Antolin}, \& {De Moortel}}]{2019Guo}
{Guo}, M., {Van Doorsselaere}, T., {Karampelas}, K., {et~al.} 2019, \apj, 870,
  55, \dodoi{10.3847/1538-4357/aaf1d0}

\bibitem[{{Karampelas} \& {Van Doorsselaere}(2020)}]{2020Karampelas}
{Karampelas}, K., \& {Van Doorsselaere}, T. 2020, \apjl, 897, L35,
  \dodoi{10.3847/2041-8213/ab9f38}

\bibitem[{{Karampelas} \& {Van Doorsselaere}(2024)}]{2024Karampelas}
---. 2024, \aap, 681, L6, \dodoi{10.1051/0004-6361/202348144}

\bibitem[{{Karampelas} {et~al.}(2017){Karampelas}, {Van Doorsselaere}, \&
  {Antolin}}]{2017Karampelas}
{Karampelas}, K., {Van Doorsselaere}, T., \& {Antolin}, P. 2017, \aap, 604,
  A130, \dodoi{10.1051/0004-6361/201730598}

\bibitem[{{Kraaikamp} {et~al.}(2023){Kraaikamp}, {Gissot}, {Stegen}, {Mampaey},
  {Verbeeck}, {Auch{\`e}re}, \& {Berghmans}}]{euidatarelease6}
{Kraaikamp}, E., {Gissot}, S., {Stegen}, K., {et~al.} 2023, SolO/EUI Data
  Release 6.0 2023-01, https://doi.org/10.24414/z818-4163

\bibitem[{Lemen {et~al.}(2012)Lemen, Title, Akin, Boerner, Chou, Drake, Duncan,
  Edwards, Friedlaender, Heyman, Hurlburt, Katz, Kushner, Levay, Lindgren,
  Mathur, McFeaters, Mitchell, Rehse, Schrijver, Springer, Stern, Tarbell,
  Wuelser, Wolfson, Yanari, Bookbinder, Cheimets, Caldwell, Deluca, Gates,
  Golub, Park, Podgorski, Bush, Scherrer, Gummin, Smith, Auker, Jerram, Pool,
  Soufli, Windt, Beardsley, Clapp, Lang, \& Waltham}]{2012Lemen}
Lemen, J.~R., Title, A.~M., Akin, D.~J., {et~al.} 2012, \solphys, 275, 17,
  \dodoi{10.1007/978-1-4614-3673-7_3}

\bibitem[{{Li} {et~al.}(2023{\natexlab{a}}){Li}, {Bai}, {Tian}, {Su}, {Hou},
  {Deng}, {Ji}, \& {Ning}}]{2023Li_travelling}
{Li}, D., {Bai}, X., {Tian}, H., {et~al.} 2023{\natexlab{a}}, \aap, 675, A169,
  \dodoi{10.1051/0004-6361/202245812}

\bibitem[{{Li} \& {Long}(2023)}]{2023Li}
{Li}, D., \& {Long}, D.~M. 2023, \apj, 944, 8, \dodoi{10.3847/1538-4357/acacf4}

\bibitem[{{Li} {et~al.}(2023{\natexlab{b}}){Li}, {Li}, {Shi}, {Su}, {Chen},
  {Yu}, {Li}, {Qiu}, {Huang}, \& {Ning}}]{2023Li_cont}
{Li}, D., {Li}, Z., {Shi}, F., {et~al.} 2023{\natexlab{b}}, \aap, 680, L15,
  \dodoi{10.1051/0004-6361/202348075}

\bibitem[{{Lim} {et~al.}(2023){Lim}, {Van Doorsselaere}, {Berghmans}, {Morton},
  {Pant}, \& {Mandal}}]{2023lim}
{Lim}, D., {Van Doorsselaere}, T., {Berghmans}, D., {et~al.} 2023, \apjl, 952,
  L15, \dodoi{10.3847/2041-8213/ace423}

\bibitem[{{Lim} {et~al.}(2024){Lim}, {Van Doorsselaere}, {Berghmans}, \&
  {Petrova}}]{2024Lim}
{Lim}, D., {Van Doorsselaere}, T., {Berghmans}, D., \& {Petrova}, E. 2024,
  \aap, 689, A16, \dodoi{10.1051/0004-6361/202450433}

\bibitem[{{Lopin} \& {Nagorny}(2024)}]{2024Lopin_Nagorny}
{Lopin}, I., \& {Nagorny}, I. 2024, \mnras, 527, 5741,
  \dodoi{10.1093/mnras/stad3527}

\bibitem[{{Mandal} {et~al.}(2021){Mandal}, {Tian}, \&
  {Peter}}]{2021Mandal_flareoscl}
{Mandal}, S., {Tian}, H., \& {Peter}, H. 2021, \aap, 652, L3,
  \dodoi{10.1051/0004-6361/202141542}

\bibitem[{{Mandal} {et~al.}(2022){Mandal}, {Chitta}, {Antolin}, {Peter},
  {Solanki}, {Auch{\`e}re}, {Berghmans}, {Zhukov}, {Teriaca}, {Cuadrado},
  {Sch{\"u}hle}, {Parenti}, {Buchlin}, {Harra}, {Verbeeck}, {Kraaikamp},
  {Long}, {Rodriguez}, {Pelouze}, {Schwanitz}, {Barczynski}, \&
  {Smith}}]{2022Mandal}
{Mandal}, S., {Chitta}, L.~P., {Antolin}, P., {et~al.} 2022, \aap, 666, L2,
  \dodoi{10.1051/0004-6361/202244403}

\bibitem[{{Morgan} \& {Druckm{\"u}ller}(2014)}]{Morgan&miller2014}
{Morgan}, H., \& {Druckm{\"u}ller}, M. 2014, \solphys, 289, 2945,
  \dodoi{10.1007/s11207-014-0523-9}

\bibitem[{{Morton} \& {McLaughlin}(2013)}]{2013Morton_McLaughlin}
{Morton}, R.~J., \& {McLaughlin}, J.~A. 2013, \aap, 553, L10,
  \dodoi{10.1051/0004-6361/201321465}

\bibitem[{{Morton} \& {McLaughlin}(2014)}]{2014Morton}
---. 2014, \apj, 789, 105, \dodoi{10.1088/0004-637X/789/2/105}

\bibitem[{{M{\"u}ller} {et~al.}(2020){M{\"u}ller}, {St. Cyr}, {Zouganelis},
  {Gilbert}, {Marsden}, {Nieves-Chinchilla}, {Antonucci}, {Auch{\`e}re},
  {Berghmans}, {Horbury}, {Howard}, {Krucker}, {Maksimovic}, {Owen}, {Rochus},
  {Rodriguez-Pacheco}, {Romoli}, {Solanki}, {Bruno}, {Carlsson}, {Fludra},
  {Harra}, {Hassler}, {Livi}, {Louarn}, {Peter}, {Sch{\"u}hle}, {Teriaca}, {del
  Toro Iniesta}, {Wimmer-Schweingruber}, {Marsch}, {Velli}, {De Groof},
  {Walsh}, \& {Williams}}]{2020Mulller}
{M{\"u}ller}, D., {St. Cyr}, O.~C., {Zouganelis}, I., {et~al.} 2020, \aap, 642,
  A1, \dodoi{10.1051/0004-6361/202038467}

\bibitem[{{Nakariakov} {et~al.}(2016){Nakariakov}, {Anfinogentov},
  {Nistic{\`o}}, \& {Lee}}]{2016Nakariakov}
{Nakariakov}, V.~M., {Anfinogentov}, S.~A., {Nistic{\`o}}, G., \& {Lee}, D.~H.
  2016, \aap, 591, L5, \dodoi{10.1051/0004-6361/201628850}

\bibitem[{{Nakariakov} \& {Ofman}(2001)}]{2001Nakariakov_Ofman}
{Nakariakov}, V.~M., \& {Ofman}, L. 2001, \aap, 372, L53,
  \dodoi{10.1051/0004-6361:20010607}

\bibitem[{{Nakariakov} {et~al.}(1999){Nakariakov}, {Ofman}, {Deluca},
  {Roberts}, \& {Davila}}]{1999Nakariakov}
{Nakariakov}, V.~M., {Ofman}, L., {Deluca}, E.~E., {Roberts}, B., \& {Davila},
  J.~M. 1999, Science, 285, 862, \dodoi{10.1126/science.285.5429.862}

\bibitem[{{Nakariakov} {et~al.}(2021){Nakariakov}, {Anfinogentov}, {Antolin},
  {Jain}, {Kolotkov}, {Kupriyanova}, {Li}, {Magyar}, {Nistic{\`o}}, {Pascoe},
  {Srivastava}, {Terradas}, {Vasheghani Farahani}, {Verth}, {Yuan}, \&
  {Zimovets}}]{2021NakariakovSSRv}
{Nakariakov}, V.~M., {Anfinogentov}, S.~A., {Antolin}, P., {et~al.} 2021, \ssr,
  217, 73, \dodoi{10.1007/s11214-021-00847-2}

\bibitem[{{Nechaeva} {et~al.}(2019){Nechaeva}, {Zimovets}, {Nakariakov}, \&
  {Goddard}}]{2019Nechaeva}
{Nechaeva}, A., {Zimovets}, I.~V., {Nakariakov}, V.~M., \& {Goddard}, C.~R.
  2019, \apjs, 241, 31, \dodoi{10.3847/1538-4365/ab0e86}

\bibitem[{{Nistic{\`o}} {et~al.}(2013){Nistic{\`o}}, {Nakariakov}, \&
  {Verwichte}}]{2013Nistico}
{Nistic{\`o}}, G., {Nakariakov}, V.~M., \& {Verwichte}, E. 2013, \aap, 552,
  A57, \dodoi{10.1051/0004-6361/201220676}

\bibitem[{{Pant} {et~al.}(2015){Pant}, {Datta}, \& {Banerjee}}]{2015Pant}
{Pant}, V., {Datta}, A., \& {Banerjee}, D. 2015, \apjl, 801, L2,
  \dodoi{10.1088/2041-8205/801/1/L2}

\bibitem[{{Pascoe} {et~al.}(2016){Pascoe}, {Goddard}, {Nistic{\`o}},
  {Anfinogentov}, \& {Nakariakov}}]{2016Pascoe}
{Pascoe}, D.~J., {Goddard}, C.~R., {Nistic{\`o}}, G., {Anfinogentov}, S., \&
  {Nakariakov}, V.~M. 2016, \aap, 589, A136,
  \dodoi{10.1051/0004-6361/201628255}

\bibitem[{{Petrova} {et~al.}(2023){Petrova}, {Magyar}, {Van Doorsselaere}, \&
  {Berghmans}}]{2023Petrova}
{Petrova}, E., {Magyar}, N., {Van Doorsselaere}, T., \& {Berghmans}, D. 2023,
  \apj, 946, 36, \dodoi{10.3847/1538-4357/acb26a}

\bibitem[{{Rochus} {et~al.}(2020){Rochus}, {Auch{\`e}re}, {Berghmans}, {Harra},
  {Schmutz}, {Sch{\"u}hle}, {Addison}, {Appourchaux}, {Aznar Cuadrado},
  {Baker}, {Barbay}, {Bates}, {BenMoussa}, {Bergmann}, {Beurthe}, {Borgo},
  {Bonte}, {Bouzit}, {Bradley}, {B{\"u}chel}, {Buchlin}, {B{\"u}chner},
  {Cab{\'e}}, {Cadiergues}, {Chaigneau}, {Chares}, {Choque Cortez}, {Coker},
  {Condamin}, {Coumar}, {Curdt}, {Cutler}, {Davies}, {Davison}, {Defise}, {Del
  Zanna}, {Delmotte}, {Delouille}, {Dolla}, {Dumesnil}, {D{\"u}rig}, {Enge},
  {Fran{\c{c}}ois}, {Fourmond}, {Gillis}, {Giordanengo}, {Gissot}, {Green},
  {Guerreiro}, {Guilbaud}, {Gyo}, {Haberreiter}, {Hafiz}, {Hailey}, {Halain},
  {Hansotte}, {Hecquet}, {Heerlein}, {Hellin}, {Hemsley}, {Hermans}, {Hervier},
  {Hochedez}, {Houbrechts}, {Ihsan}, {Jacques}, {J{\'e}r{\^o}me}, {Jones},
  {Kahle}, {Kennedy}, {Klaproth}, {Kolleck}, {Koller}, {Kotsialos},
  {Kraaikamp}, {Langer}, {Lawrenson}, {Le Clech'}, {Lenaerts}, {Liebecq},
  {Linder}, {Long}, {Mampaey}, {Markiewicz-Innes}, {Marquet}, {Marsch},
  {Matthews}, {Mazy}, {Mazzoli}, {Meining}, {Meltchakov}, {Mercier}, {Meyer},
  {Monecke}, {Monfort}, {Morinaud}, {Moron}, {Mountney}, {M{\"u}ller},
  {Nicula}, {Parenti}, {Peter}, {Pfiffner}, {Philippon}, {Phillips},
  {Plesseria}, {Pylyser}, {Rabecki}, {Ravet-Krill}, {Rebellato}, {Renotte},
  {Rodriguez}, {Roose}, {Rosin}, {Rossi}, {Roth}, {Rouesnel}, {Roulliay},
  {Rousseau}, {Ruane}, {Scanlan}, {Schlatter}, {Seaton}, {Silliman}, {Smit},
  {Smith}, {Solanki}, {Spescha}, {Spencer}, {Stegen}, {Stockman}, {Szwec},
  {Tamiatto}, {Tandy}, {Teriaca}, {Theobald}, {Tychon}, {van Driel-Gesztelyi},
  {Verbeeck}, {Vial}, {Werner}, {West}, {Westwood}, {Wiegelmann}, {Willis},
  {Winter}, {Zerr}, {Zhang}, \& {Zhukov}}]{2020rochus}
{Rochus}, P., {Auch{\`e}re}, F., {Berghmans}, D., {et~al.} 2020, \aap, 642, A8,
  \dodoi{10.1051/0004-6361/201936663}

\bibitem[{{Ruderman} \& {Petrukhin}(2021)}]{2021Ruderman}
{Ruderman}, M.~S., \& {Petrukhin}, N.~S. 2021, \mnras, 501, 3017,
  \dodoi{10.1093/mnras/staa3816}

\bibitem[{{Schrijver} {et~al.}(1999){Schrijver}, {Title}, {Berger}, {Fletcher},
  {Hurlburt}, {Nightingale}, {Shine}, {Tarbell}, {Wolfson}, {Golub},
  {Bookbinder}, {Deluca}, {McMullen}, {Warren}, {Kankelborg}, {Handy}, \& {de
  Pontieu}}]{1999Schrijver}
{Schrijver}, C.~J., {Title}, A.~M., {Berger}, T.~E., {et~al.} 1999, \solphys,
  187, 261, \dodoi{10.1023/A:1005194519642}

\bibitem[{{Shi} {et~al.}(2021){Shi}, {Van Doorsselaere}, {Guo}, {Karampelas},
  {Li}, \& {Antolin}}]{2021Shi}
{Shi}, M., {Van Doorsselaere}, T., {Guo}, M., {et~al.} 2021, \apj, 908, 233,
  \dodoi{10.3847/1538-4357/abda54}

\bibitem[{{Shrivastav} {et~al.}(2024{\natexlab{a}}){Shrivastav}, {Pant}, \&
  {Antolin}}]{2024Shrivastav_coronal_rain}
{Shrivastav}, A.~K., {Pant}, V., \& {Antolin}, P. 2024{\natexlab{a}}, \aap,
  689, A295, \dodoi{10.1051/0004-6361/202449677}

\bibitem[{{Shrivastav} {et~al.}(2024{\natexlab{b}}){Shrivastav}, {Pant},
  {Berghmans}, {Zhukov}, {Van Doorsselaere}, {Petrova}, {Banerjee}, {Lim}, \&
  {Verbeeck}}]{2023Shrivastav}
{Shrivastav}, A.~K., {Pant}, V., {Berghmans}, D., {et~al.} 2024{\natexlab{b}},
  \aap, 685, A36, \dodoi{10.1051/0004-6361/202346670}

\bibitem[{{Tian} {et~al.}(2012){Tian}, {McIntosh}, {Wang}, {Ofman}, {De
  Pontieu}, {Innes}, \& {Peter}}]{2012Tian}
{Tian}, H., {McIntosh}, S.~W., {Wang}, T., {et~al.} 2012, \apj, 759, 144,
  \dodoi{10.1088/0004-637X/759/2/144}

\bibitem[{{Van Doorsselaere} {et~al.}(2020){Van Doorsselaere}, {Srivastava},
  {Antolin}, {Magyar}, {Vasheghani Farahani}, {Tian}, {Kolotkov}, {Ofman},
  {Guo}, {Arregui}, {De Moortel}, \& {Pascoe}}]{2020tom}
{Van Doorsselaere}, T., {Srivastava}, A.~K., {Antolin}, P., {et~al.} 2020,
  \ssr, 216, 140, \dodoi{10.1007/s11214-020-00770-y}

\bibitem[{{Wang} {et~al.}(2012){Wang}, {Ofman}, {Davila}, \& {Su}}]{2012Wang}
{Wang}, T., {Ofman}, L., {Davila}, J.~M., \& {Su}, Y. 2012, \apjl, 751, L27,
  \dodoi{10.1088/2041-8205/751/2/L27}

\bibitem[{{Yuan} {et~al.}(2023){Yuan}, {Fu}, {Cao}, {Ku{\'z}ma}, {Geeraerts},
  {Trelles Arjona}, {Murawski}, {Van Doorsselaere}, {Srivastava}, {Miao},
  {Feng}, {Feng}, {Quintero Noda}, {Ruiz Cobo}, \& {Su}}]{2023Yuan_natas}
{Yuan}, D., {Fu}, L., {Cao}, W., {et~al.} 2023, Nature Astronomy, 7, 856,
  \dodoi{10.1038/s41550-023-01973-3}

\bibitem[{{Zhang} {et~al.}(2022){Zhang}, {Chen}, {Li}, {Lu}, \&
  {Li}}]{2022Zhang}
{Zhang}, Q.~M., {Chen}, J.~L., {Li}, S.~T., {Lu}, L., \& {Li}, D. 2022,
  \solphys, 297, 18, \dodoi{10.1007/s11207-022-01952-3}

\bibitem[{{Zhong} {et~al.}(2021){Zhong}, {Duckenfield}, {Nakariakov}, \&
  {Anfinogentov}}]{2021Zhong_motion}
{Zhong}, S., {Duckenfield}, T.~J., {Nakariakov}, V.~M., \& {Anfinogentov},
  S.~A. 2021, \solphys, 296, 135, \dodoi{10.1007/s11207-021-01870-w}

\bibitem[{{Zhong} {et~al.}(2022{\natexlab{a}}){Zhong}, {Nakariakov},
  {Kolotkov}, \& {Anfinogentov}}]{2022Zhonga}
{Zhong}, S., {Nakariakov}, V.~M., {Kolotkov}, D.~Y., \& {Anfinogentov}, S.~A.
  2022{\natexlab{a}}, \mnras, 513, 1834, \dodoi{10.1093/mnras/stac1014}

\bibitem[{{Zhong} {et~al.}(2022{\natexlab{b}}){Zhong}, {Nakariakov},
  {Kolotkov}, {Verbeeck}, \& {Berghmans}}]{2022Zhong}
{Zhong}, S., {Nakariakov}, V.~M., {Kolotkov}, D.~Y., {Verbeeck}, C., \&
  {Berghmans}, D. 2022{\natexlab{b}}, \mnras, 516, 5989,
  \dodoi{10.1093/mnras/stac2545}

\bibitem[{{Zhong} {et~al.}(2023){Zhong}, {Nakariakov}, {Miao}, {Fu}, \&
  {Yuan}}]{2023Zhong}
{Zhong}, S., {Nakariakov}, V.~M., {Miao}, Y., {Fu}, L., \& {Yuan}, D. 2023,
  arXiv e-prints, arXiv:2308.05479, \dodoi{10.48550/arXiv.2308.05479}

\end{thebibliography}
\bibliographystyle{aasjournal}

\begin{appendices}

\section{Appendix A} \label{appendix}

\renewcommand{\arraystretch}{1.26}

\setlength{\tabcolsep}{10pt}

\begin{longtable}{cccccccc} 

 \caption{Details of estimated loop length and parameters of oscillations} \label{table2} \\

\hline

No . & L (Mm) & P (s) & A (km) & V (km/s) & C$_{k}$ (km/s) & B (G) & Mean Loop Width (Mm) \\ 
\hline
\endfirsthead

\caption*{Details of estimated loop length and parameters of oscillations } \\ 

\hline
No . & L (Mm) & P (s) & A (km) & V (km/s) & C$_{k}$ (km/s) & B (G) & Mean Loop Width (Mm)  \\ \hline
\endhead
\hline
\multicolumn{3}{r}{Continued on next page} \\ 
\endfoot
\hline
\endlastfoot

        1 & 24.5 & 252 ± 5 & 176 ± 12 & 4.4 ± 0.3 & 194 & 2.6 & 0.5 \\ \hline
        2 & 29.4 & 31 ± 2 & 50 ± 10 & 10.1 ± 2.1 & 1897 & 25.3 & 0.2 \\ \hline
        3 & 37.1 & 31 ± 1 & 86 ± 10 & 17.4 ± 2.1 & 2394 & 31.9 & 0.3 \\ \hline
        4 & 7.7 & 95 ± 2 & 68 ± 8 & 4.5 ± 0.5 & 162 & 2.2 & 0.6 \\ \hline
        5 & 8.9 & 165 ± 6 & 128 ± 15 & 4.9 ± 0.6 & 108 & 1.4 & 0.3 \\ \hline
        6 & 17.4 & 50 ± 3 & 74 ± 14 & 9.3 ± 1.8 & 696 & 9.3 & 0.3 \\ \hline
        7 & 25.1 & 147 ± 4 & 85 ± 9 & 3.6 ± 0.4 & 341 & 4.6 & 0.4 \\ \hline
        8 & 25.9 & 42 ± 2 & 58 ± 13 & 8.7 ± 2 & 1233 & 16.4 & 1.1 \\ \hline
        9 & 17.9 & 169 ± 12 & 58 ± 13 & 2.2 ± 0.5 & 212 & 2.8 & 0.3 \\ \hline
        10 & 28.9 & 25 ± 0 & 60 ± 4 & 15.1 ± 1 & 2312 & 30.8 & 1.1 \\ \hline
        11 & 24.5 & 103 ± 8 & 137 ± 16 & 8.4 ± 1.2 & 476 & 6.3 & 0.3 \\ \hline
        12 & 22.2 & 59 ± 3 & 85 ± 18 & 9.1 ± 2 & 753 & 10 & 1 \\ \hline
        13 & 42.5 & 308 ± 17 & 355 ± 50 & 7.2 ± 1.1 & 276 & 3.7 & 0.5 \\ \hline
        14 & 12.2 & 421 ± 16 & 408 ± 27 & 6.1 ± 0.5 & 58 & 0.8 & 1.1 \\ \hline
        15 & 16.3 & 165 ± 10 & 121 ± 21 & 4.6 ± 0.8 & 198 & 2.6 & 2.3 \\ \hline
        16 & 19.8 & 188 ± 11 & 52 ± 10 & 1.7 ± 0.3 & 211 & 2.8 & 0.3 \\ \hline
        17 & 12.6 & 23 ± 1 & 45 ± 14 & 12.3 ± 3.9 & 1096 & 14.6 & 1.6 \\ \hline
        18 & 13.2 & 157 ± 3 & 39 ± 3 & 1.6 ± 0.1 & 168 & 2.2 & 0.3 \\ \hline
        19 & 17.2 & 63 ± 8 & 53 ± 35 & 5.3 ± 3.6 & 546 & 7.3 & 0.4 \\ \hline
        20 & 17.2 & 75 ± 5 & 45 ± 7 & 3.8 ± 0.6 & 459 & 6.1 & 0.4 \\ \hline
        21 & 14.3 & 89 ± 5 & 37 ± 8 & 2.6 ± 0.6 & 321 & 4.3 & 0.3 \\ \hline
        22 & 10 & 262 ± 2 & 189 ± 6 & 4.5 ± 0.1 & 76 & 1 & 1.1 \\ \hline
        23 & 16.8 & 168 ± 3 & 208 ± 7 & 7.8 ± 0.3 & 200 & 2.7 & 0.3 \\ \hline
        24 & 15 & 88 ± 2 & 85 ± 7 & 6.1 ± 0.5 & 341 & 4.5 & 0.3 \\ \hline
        25 & 15.7 & 25 ± 1 & 94 ± 14 & 23.6 ± 3.6 & 1256 & 16.7 & 0.4 \\ \hline
        26 & 17.2 & 66 ± 6 & 87 ± 28 & 8.3 ± 2.8 & 521 & 6.9 & 0.2 \\ \hline
        27 & 12.8 & 99 ± 2 & 283 ± 20 & 18 ± 1.3 & 259 & 3.4 & 0.5 \\ \hline
        28 & 13.5 & 79 ± 1 & 95 ± 3 & 7.6 ± 0.3 & 342 & 4.6 & 1.7 \\ \hline
        29 & 12.8 & 359 ± 13 & 191 ± 14 & 3.3 ± 0.3 & 71 & 1 & 0.4 \\ \hline
        30 & 13.3 & 103 ± 2 & 268 ± 16 & 16.3 ± 1 & 258 & 3.4 & 1.5 \\ \hline
        31 & 31.4 & 36 ± 1 & 54 ± 6 & 9.4 ± 1.1 & 1744 & 23.3 & 0.3 \\ \hline
        32 & 14.7 & 113 ± 6 & 82 ± 8 & 4.6 ± 0.5 & 260 & 3.5 & 0.2 \\ \hline
        33 & 28.5 & 81 ± 4 & 40 ± 6 & 3.1 ± 0.5 & 704 & 9.4 & 0.3 \\ \hline
        34 & 16.4 & 256 ± 6 & 100 ± 6 & 2.5 ± 0.2 & 128 & 1.7 & 1.6 \\ \hline
        35 & 14.7 & 70 ± 5 & 57 ± 13 & 5.1 ± 1.2 & 420 & 5.6 & 0.2 \\ \hline
        36 & 19.9 & 104 ± 10 & 57 ± 18 & 3.4 ± 1.1 & 383 & 5.1 & 1.7 \\ \hline
        37 & 20.5 & 434 ± 6 & 186 ± 16 & 2.7 ± 0.2 & 94 & 1.3 & 2 \\ \hline
        38 & 26 & 169 ± 11 & 86 ± 14 & 3.2 ± 0.6 & 308 & 4.1 & 0.3 \\ \hline
        39 & 21.4 & 226 ± 4 & 196 ± 10 & 5.4 ± 0.3 & 189 & 2.5 & 1.9 \\ \hline
        40 & 21.2 & 191 ± 6 & 118 ± 15 & 3.9 ± 0.5 & 222 & 3 & 0.4 \\ \hline
        41 & 21.3 & 149 ± 17 & 81 ± 17 & 3.4 ± 0.8 & 286 & 3.8 & 3.2 \\ \hline
        42 & 14.7 & 73 ± 3 & 204 ± 28 & 17.6 ± 2.5 & 403 & 5.4 & 0.3 \\ \hline
        43 & 28.8 & 27 ± 1 & 56 ± 11 & 13 ± 2.6 & 2133 & 28.4 & 0.2 \\ \hline
        44 & 4.5 & 265 ± 4 & 110 ± 7 & 2.6 ± 0.2 & 34 & 0.5 & 0.3 \\ \hline
        45 & 13.5 & 155 ± 8 & 148 ± 17 & 6 ± 0.8 & 174 & 2.3 & 0.2 \\ \hline
        46 & 14.6 & 91 ± 4 & 60 ± 7 & 4.1 ± 0.5 & 321 & 4.3 & 0.3 \\ \hline
        47 & 21.7 & 442 ± 3 & 325 ± 6 & 4.6 ± 0.1 & 98 & 1.3 & 0.4 \\ \hline
        48 & 16.9 & 33 ± 1 & 62 ± 10 & 11.8 ± 1.9 & 1024 & 13.7 & 1.7 \\ \hline
        49 & 16.2 & 72 ± 1 & 269 ± 9 & 23.5 ± 0.9 & 450 & 6 & 0.4 \\ \hline
        50 & 29.3 & 35 ± 1 & 35 ± 5 & 6.3 ± 0.9 & 1674 & 22.3 & 0.3 \\ \hline
        51 & 19.8 & 102 ± 2 & 261 ± 9 & 16.1 ± 0.6 & 388 & 5.2 & 0.5 \\ \hline
        52 & 30.3 & 111 ± 13 & 52 ± 20 & 2.9 ± 1.2 & 546 & 7.3 & 0.3 \\ \hline
        53 & 49.1 & 167 ± 1 & 162 ± 3 & 6.1 ± 0.1 & 588 & 7.8 & 0.3 \\ \hline
        54 & 14.6 & 85 ± 7 & 46 ± 7 & 3.4 ± 0.6 & 344 & 4.6 & 0.2 \\ \hline
        55 & 10 & 234 ± 19 & 120 ± 21 & 3.2 ± 0.6 & 85 & 1.1 & 0.5 \\ \hline
        56 & 14.6 & 140 ± 12 & 35 ± 5 & 1.6 ± 0.3 & 209 & 2.8 & 0.5 \\ \hline
        57 & 10.4 & 86 ± 2 & 43 ± 4 & 3.1 ± 0.3 & 242 & 3.2 & 0.2 \\ \hline
        58 & 14.6 & 93 ± 5 & 36 ± 7 & 2.4 ± 0.5 & 314 & 4.2 & 0.7 \\ \hline
        59 & 14.6 & 46 ± 1 & 30 ± 4 & 4.1 ± 0.6 & 635 & 8.5 & 0.2 \\ \hline
        60 & 15 & 101 ± 2 & 43 ± 4 & 2.7 ± 0.3 & 297 & 4 & 0.2 \\ \hline
        61 & 17.6 & 31 ± 1 & 67 ± 9 & 13.6 ± 1.9 & 1135 & 15.1 & 2.2 \\ \hline
        62 & 30.4 & 347 ± 19 & 262 ± 39 & 4.7 ± 0.8 & 175 & 2.3 & 0.9 \\ \hline
        63 & 12.9 & 25 ± 1 & 66 ± 9 & 16.6 ± 2.4 & 1032 & 13.8 & 3.3 \\ \hline
        64 & 8.1 & 179 ± 7 & 129 ± 16 & 4.5 ± 0.6 & 91 & 1.2 & 0.3 \\ \hline
        65 & 18.7 & 313 ± 14 & 157 ± 14 & 3.2 ± 0.3 & 119 & 1.6 & 1.6 \\ \hline
        66 & 7.6 & 205 ± 11 & 85 ± 8 & 2.6 ± 0.3 & 74 & 1 & 0.9 \\ \hline
        67 & 19.8 & 219 ± 8 & 78 ± 11 & 2.2 ± 0.3 & 181 & 2.4 & 2.5 \\ \hline
        68 & 16.2 & 126 ± 4 & 188 ± 22 & 9.4 ± 1.1 & 257 & 3.4 & 0.4 \\ \hline
        69 & 16.9 & 120 ± 12 & 76 ± 17 & 4 ± 1 & 282 & 3.8 & 0.9 \\ \hline
        70 & 21 & 179 ± 5 & 34 ± 6 & 1.2 ± 0.2 & 235 & 3.1 & 0.7 \\ \hline
        71 & 16.9 & 467 ± 7 & 108 ± 7 & 1.5 ± 0.1 & 72 & 1 & 1.8 \\ \hline
        72 & 13.9 & 95 ± 20 & 54 ± 25 & 3.6 ± 1.8 & 293 & 3.9 & 1.9 \\ \hline
        73 & 19.1 & 222 ± 24 & 89 ± 20 & 2.5 ± 0.6 & 172 & 2.3 & 0.8 \\ \hline
        74 & 30 & 149 ± 16 & 67 ± 24 & 2.8 ± 1.1 & 403 & 5.4 & 0.3 \\ \hline
        75 & 23 & 205 ± 14 & 96 ± 25 & 2.9 ± 0.8 & 224 & 3 & 1 \\ \hline
        76 & 8.2 & 205 ± 8 & 155 ± 20 & 4.8 ± 0.6 & 80 & 1.1 & 2.6 \\ \hline
        77 & 12.6 & 224 ± 13 & 85 ± 20 & 2.4 ± 0.6 & 112 & 1.5 & 1.1 \\ \hline
        78 & 21 & 404 ± 9 & 161 ± 12 & 2.5 ± 0.2 & 104 & 1.4 & 0.4 \\ \hline
        79 & 15 & 296 ± 13 & 129 ± 13 & 2.7 ± 0.3 & 101 & 1.4 & 0.9 \\ \hline
        80 & 20.8 & 308 ± 11 & 57 ± 7 & 1.2 ± 0.1 & 135 & 1.8 & 0.4 \\ \hline
        81 & 24.2 & 275 ± 19 & 103 ± 18 & 2.4 ± 0.4 & 176 & 2.3 & 0.5 \\ \hline
        82 & 9.4 & 40 ± 2 & 39 ± 6 & 6.1 ± 1 & 470 & 6.3 & 0.7 \\ \hline
        83 & 11.7 & 148 ± 6 & 42 ± 6 & 1.8 ± 0.3 & 158 & 2.1 & 1.7 \\ \hline
        84 & 14.6 & 224 ± 6 & 69 ± 7 & 1.9 ± 0.2 & 130 & 1.7 & 0.5 \\ \hline
        85 & 11.8 & 24 ± 2 & 36 ± 10 & 9.4 ± 2.7 & 983 & 13.1 & 0.2 \\ \hline
        86 & 11.5 & 37 ± 1 & 44 ± 7 & 7.5 ± 1.2 & 622 & 8.3 & 0.5 \\ \hline
        87 & 11.7 & 112 ± 3 & 103 ± 7 & 5.8 ± 0.4 & 209 & 2.8 & 0.3 \\ \hline
        88 & 7.7 & 40 ± 2 & 53 ± 12 & 8.3 ± 1.9 & 385 & 5.1 & 0.3 \\ \hline
        89 & 25.4 & 26 ± 1 & 50 ± 18 & 12.1 ± 4.4 & 1954 & 26 & 0.2 \\ \hline
        90 & 32.2 & 127 ± 5 & 97 ± 12 & 4.8 ± 0.6 & 507 & 6.8 & 0.6 \\ \hline
        91 & 42.3 & 169 ± 6 & 163 ± 19 & 6.1 ± 0.7 & 501 & 6.7 & 1.3 \\ \hline
        92 & 42.3 & 182 ± 8 & 102 ± 12 & 3.5 ± 0.4 & 465 & 6.2 & 0.5 \\ \hline
        93 & 26.3 & 247 ± 11 & 53 ± 7 & 1.3 ± 0.2 & 213 & 2.8 & 0.7 \\ \hline
        94 & 40.1 & 243 ± 6 & 77 ± 7 & 2 ± 0.2 & 330 & 4.4 & 0.6 \\ \hline
        95 & 42.2 & 204 ± 6 & 64 ± 9 & 2 ± 0.3 & 414 & 5.5 & 0.4 \\ \hline
        96 & 35.2 & 250 ± 10 & 61 ± 8 & 1.5 ± 0.2 & 282 & 3.8 & 0.5 \\ \hline
        97 & 11.4 & 23 ± 2 & 46 ± 18 & 12.6 ± 5 & 991 & 13.2 & 1.2 \\ \hline
        98 & 12 & 34 ± 1 & 50 ± 7 & 9.2 ± 1.3 & 706 & 9.4 & 0.8 \\ \hline
        99 & 10.8 & 220 ± 5 & 163 ± 8 & 4.7 ± 0.3 & 98 & 1.3 & 0.4 \\ \hline
        100 & 8.5 & 68 ± 3 & 41 ± 6 & 3.8 ± 0.6 & 250 & 3.3 & 0.2 \\ \hline
        101 & 9.1 & 230 ± 6 & 102 ± 7 & 2.8 ± 0.2 & 79 & 1.1 & 0.8 \\ \hline
        102 & 12.8 & 23 ± 1 & 31 ± 3 & 8.5 ± 0.9 & 1113 & 14.8 & 0.2 \\ \hline
        103 & 10.3 & 36 ± 3 & 34 ± 10 & 5.9 ± 1.8 & 572 & 7.6 & 1 \\ \hline
        104 & 21.5 & 102 ± 5 & 35 ± 10 & 2.2 ± 0.6 & 422 & 5.6 & 0.8 \\ \hline
        105 & 4.1 & 339 ± 7 & 78 ± 5 & 1.4 ± 0.1 & 24 & 0.3 & 0.4 \\ \hline

\end{longtable}

\begin{figure*}[!ht]
\centering
\includegraphics[width=0.6\textwidth,clip,trim=0cm 0cm 0cm 0cm]{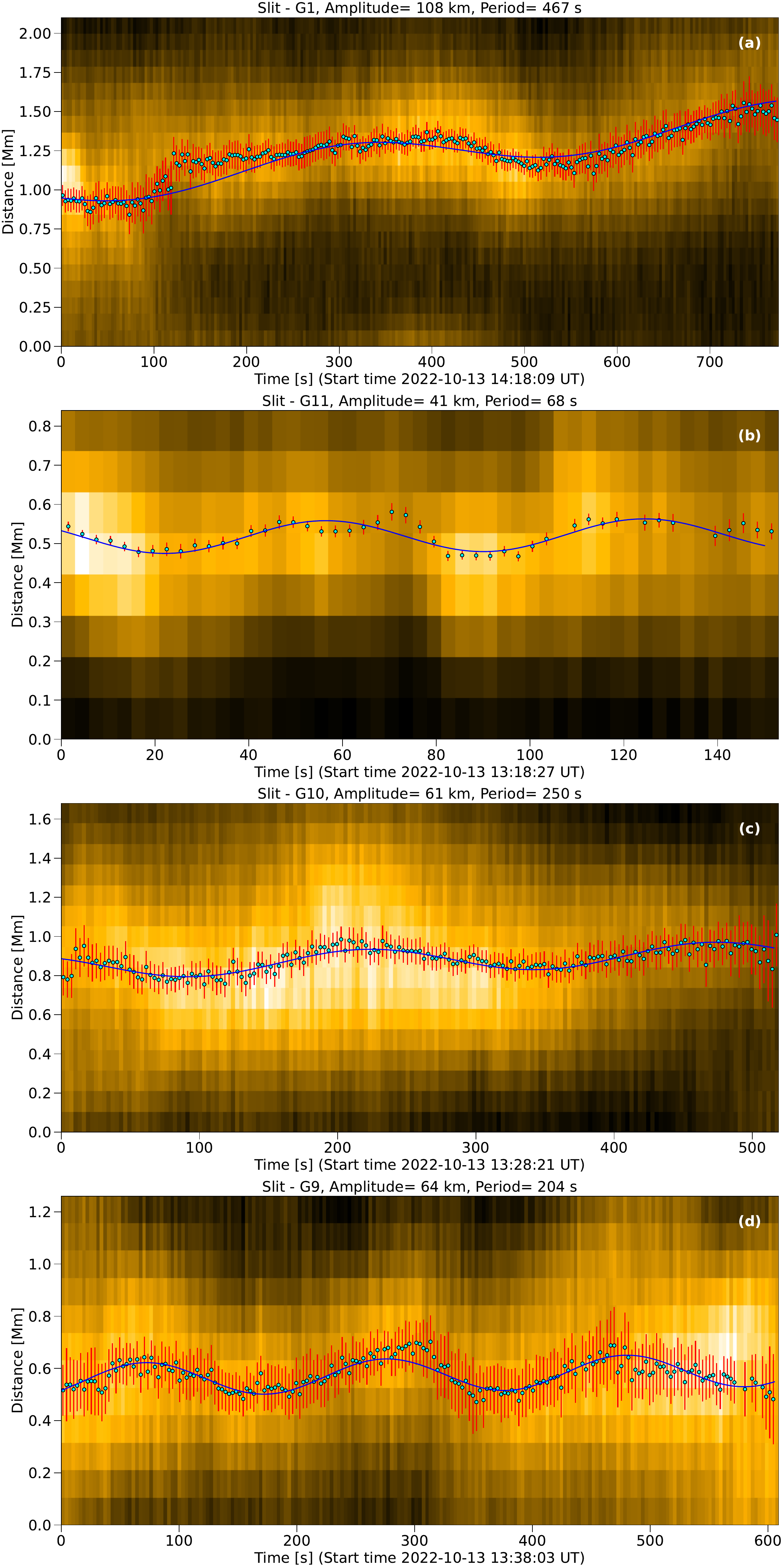}
\caption{The columns display examples of $x-t$ maps from slits G1, G11, G10, and G9, as shown in Figure \ref{fig:context_d2}. The cyan points denote the loop positions obtained from Gaussian fitting, while the blue curve represents the fitted oscillation profile.  }
\label{fig:xt-maps_appendix1}
\end{figure*}

\begin{figure*}[!ht]
\centering
\includegraphics[width=0.6\textwidth,clip,trim=0cm 0cm 0cm 0cm]{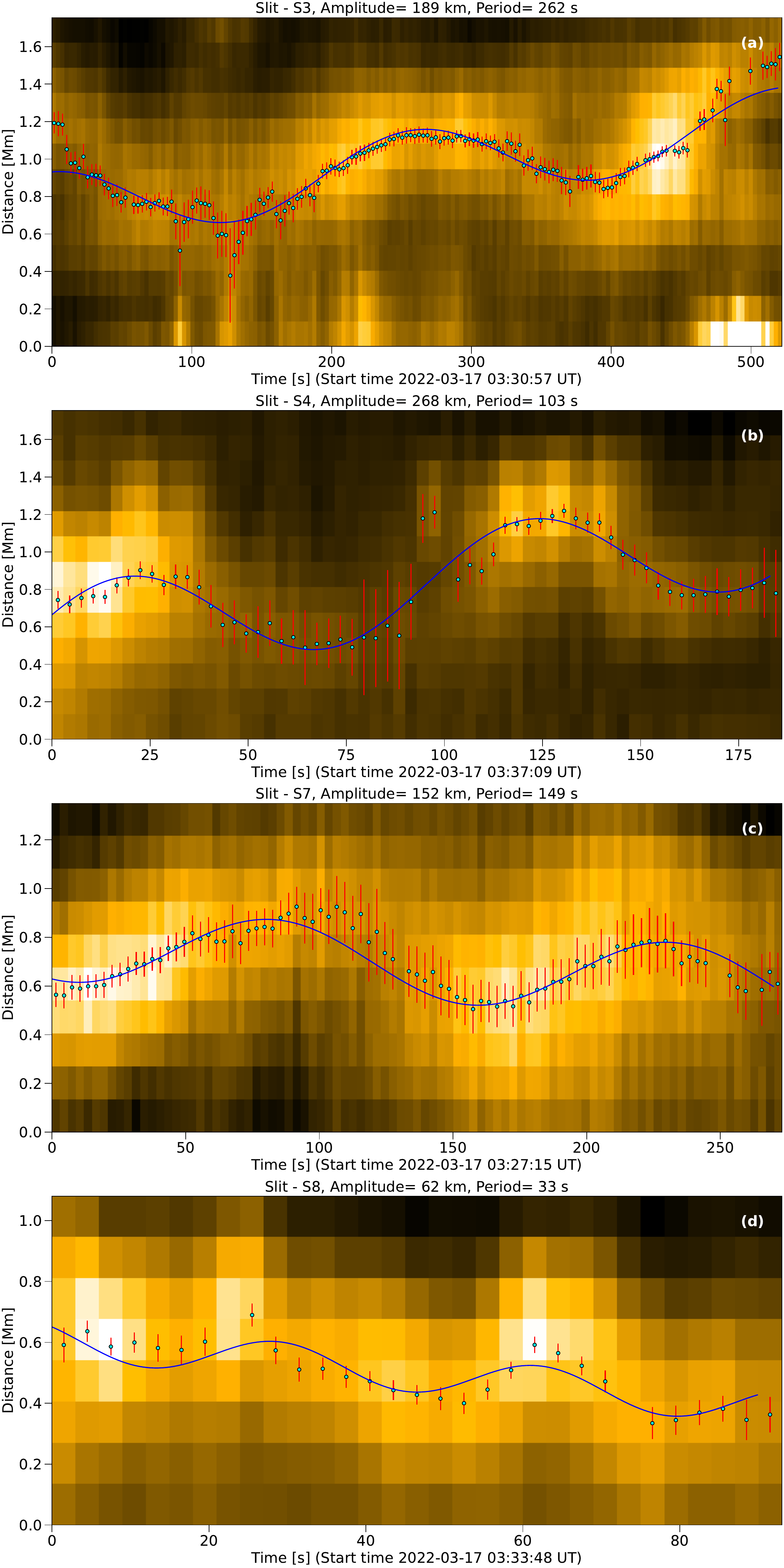}
\caption{Similar to Figure \ref{fig:xt-maps_appendix1} but $x-t$ maps are generated from slits S3, S4, S7, and S8.}
\label{fig:xt-maps_appendix2}
\end{figure*}

\end{appendices}
 
\end{document}